\PassOptionsToPackage{table}{xcolor}
\documentclass[letterpaper,twocolumn,10pt]{article}
\usepackage{hegroup}

\usepackage{tikz}
\usepackage{graphicx}
\usepackage{amsmath}
\usepackage[table,xcdraw]{xcolor}
\usepackage{tcolorbox}
\tcbuselibrary{most,listings,breakable}

\usepackage{hyperref}
\usepackage{cleveref}
\usepackage{multirow}
\usepackage{xspace}

\newtheorem{proposition}{Proposition}

\usepackage{amsfonts}
\newcommand{\mypara}[1]{\smallskip\noindent{\bf {#1}.}\xspace}
\usepackage{enumitem}
 \usepackage{booktabs}
\usepackage[normalem]{ulem}
\useunder{\uline}{\ul}{}

\definecolor{maincolor}{HTML}{1f77b4} 
\definecolor{highlight}{HTML}{4169E1} 

\newtcolorbox{cotbox}[1][]{
    colback=maincolor!10,
    colframe=maincolor,
    width=\columnwidth,
    fonttitle=\bfseries,
    coltitle=white,
    arc=1mm,
    auto outer arc,
    left=4pt,
    right=4pt,
    breakable,
    title=#1,
}

\begin{document}

\date{}

\title{A Watermark for Vision-Language-Action and World Action Models}

\author{
Yule Liu\textsuperscript{1,2}  \ \ \ 
Shuai Liu\textsuperscript{3} \ \ \
Jiaheng Wei\textsuperscript{1} \ \ \
Xinlei He\textsuperscript{2}\textsuperscript{\textdagger} \ \ \ 
\\
\\
\textsuperscript{1}\textit{Hong Kong University of Science and Technology (Guangzhou)} \ \ \  \\
\textsuperscript{2}\textit{Wuhan University} \ \ \ 
\textsuperscript{3}\textit{Xi'an Jiao Tong University} 
}

\maketitle
\footnotetext{\textsuperscript{\textdagger}Corresponding author: Xinlei He (\href{mailto:xinleihe@hkust-gz.edu.cn}{xinlei.he@whu.edu.cn}).}

\begin{abstract}

Vision-language-action (VLA) models and world-action models (WAM) are the generative models now driving general-purpose robot control, turning raw camera input directly into motor commands.
They are increasingly deployed as black-box services, where a partner runs the policy through an interface while the owner keeps the weights private.
Training such a model takes proprietary data and heavy computational power, making the deployed model itself a valuable intellectual property.
To protect the model ownership, one standard tool is watermarking.
Recent research has developed two types of watermarks for robot policies, i.e., backdoor-based and output-perturbation watermarks.
However, the backdoor-based watermark cannot be extended to the multi-user identification scenario, while the output-perturbation watermark makes the watermark signal distributed in a narrow frequency band of the action output, which can be easily detected by the adversary.

To address this, we propose the \emph{keyed latent-provenance verification} method, which watermarks the policy through the seed of the Gaussian noise vector that the models draw before generation.
At the injection stage, the owner swaps this seed for a keyed one with the same distribution as ordinary noise, so the fingerprinted actions are statistically identical to those of an ordinary run and an adversary watching the output finds no signal to detect or remove.
At the verification stage, the owner runs the suspect model under authorized access and records the action channels the robot executes, a partial and possibly post-processed view of the policy's output.
From this view, the verifier recovers the seed by gradient-based maximum a posteriori (MAP) optimization, tests it for the secret key to score each rollout, and aggregates these scores into a single decision on whether the suspect model belongs to the owner.

We evaluate the method on two representative models across two robot suites.
The experiments cover detection of the fingerprint, identification of which of several keys a suspect carries, robustness to a range of attacks, and an analysis of why the design works.
Across both models, the fingerprint can be detected reliably with little change to task performance, and it remains detectable under output-side removal attacks and weight-level edits.
The code is available in \url{https://github.com/Y-L-LIU/keyed-latent-watermarking-vla}.
\end{abstract}

\section{Introduction}

Robotic control is moving from task-specific controllers to generative policies trained on robot demonstrations, visual data, and web-scale language supervision~\cite{black2024pi0,pi05_2025,openvla,groot_n1,gemini_robotics}.
Two model families drive this shift: vision-language-action (VLA) models map camera images and a language instruction directly to motor actions~\cite{black2024pi0,pi05_2025,intelligence2026pi}, while world-action models (WAMs) first predict how the scene will change and then infer the action that produces that change~\cite{lingbotva2026, ye2026world}.
Although they differ internally, both expose the same deployment interface: run the policy on the observation and task command, and the robot receives an action chunk.
This action interface is also what a verifier can observe after deployment.

Research policies are often released with open weights or code~\cite{openvla,groot_n1}, but production policies are usually private derivatives.
To reach better performance in real-world scenarios, a provider adapts a base policy with proprietary robot demonstrations and customer-specific task data.
The production version may also use a larger model, more computational resources, or heavier serving infrastructure than a public research release.
These adaptations make the deployed policy valuable intellectual property.
The provider may serve it through an API or deliver it as a packaged service that a partner runs on site~\cite{skild_brain,intelligence2025pi06vlalearnsexperience,intelligence2026pi}.
The weights can remain hidden in both cases, but a partner who can run the service can still query it, redeploy it, or dispute where a derived system came from.
The ownership verification question is: \textit{from the action commands that reach the robot, can the provider detect a suspect service that is reusing the protected deployment to avoid training its own policy?}
\begin{figure*}
    \centering
    \includegraphics[width=1\linewidth]{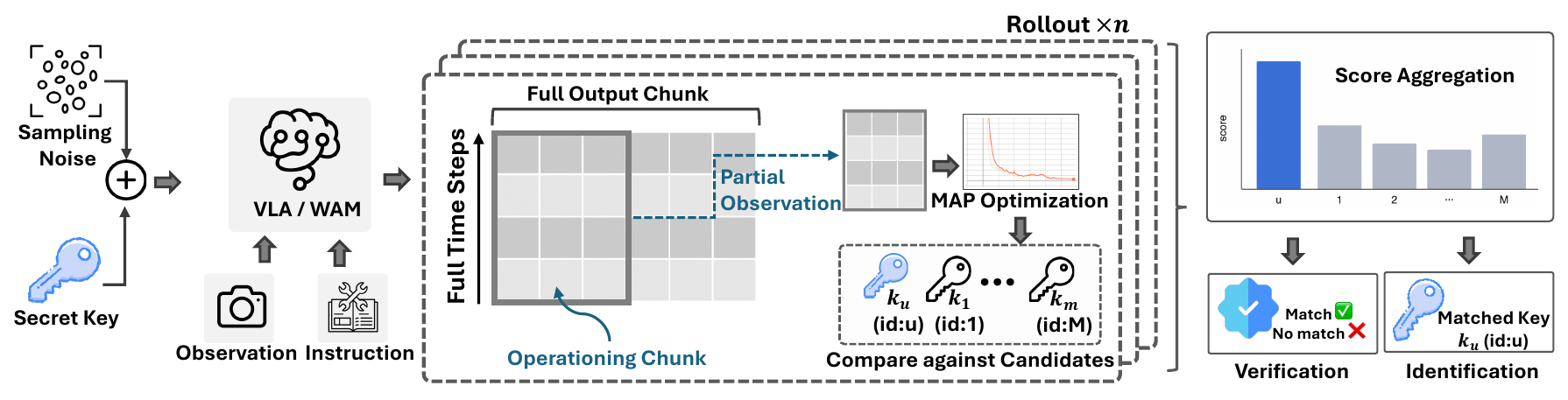}
    \caption{Overview of our method.}
    \label{fig:teaser}
\end{figure*}

Watermarking is the standard tool for ownership protection, but deployed robot policies need a mark that makes low-cost evasion insufficient.
Existing methods can be categorized into backdoor-based or output-side watermarks. 
The backdoor-based watermark changes the weights so that a secret trigger makes the policy emit a telltale action~\cite{guardvla}.
It requires weight edits and trigger calibration.
Multi-customer deployments also need separate trigger behaviors.
Later fine-tuning or behavior cloning on ordinary task outputs can weaken the rare trigger behavior.
The output-side watermark marks a stochastic policy's exploration noise~\cite{conoco}.
It shapes that noise into a secret frequency band and reads the band back from the robot's trajectory.
That signal is easy for the owner to score, but its visibility gives an adversary a cheap removal handle: spectral analysis locates the band, and a narrow filter removes it with little cost to task performance (\Cref{sec:eval-design-injection}).
Latent-noise watermarks for image generators, including Tree-Ring, Gaussian Shading, and RingID, embed ownership evidence in the keyed generation process~\cite{wen2023treering,yang2024gaussianshading,ci2024ringid}.
They suggest a latent-side route for robot policies, but the audit surface is different: the owner can inspect only the action commands that reach the robot.

Inspired by these latent-noise methods, our method has two stages (see \Cref{fig:teaser}).
In the injection stage, the challenge is to make the policy carry a key-dependent signal without changing the sampler's noise law or exposing a simple removal target in the action stream.
We therefore mark the random seed that the generator already draws before producing each action chunk.
This seed is shared by both model families: the VLA can map it directly to an action chunk, while the WAM can use it to generate an intermediate future scene from which actions are inferred.
At deployment, the owner mixes the ordinary Gaussian seed with a keyed Gaussian reference.
The mixed seed still has the same Gaussian distribution as an ordinary seed, so individual sampler draws look like ordinary noise.
The key appears only as a hidden correlation between the seed and the owner's reference.
The generator spreads that correlation through the action trajectory instead of adding an explicit output signal.

In the verification stage, the challenge is the reverse one: the owner must recover evidence for that hidden correlation without seeing the seed.
During an audit, the owner may only record the action commands that reach the robot's actuators.
These commands are a partial and altered view of the generator output.
Some raw action channels may never reach the robot, and the deployment may clip, smooth, delay, or otherwise change the channels that do.
For each marked chunk, the verifier searches for a latent seed that would make the owner's generator reproduce the observed action commands while staying likely under the Gaussian prior.
We frame this search as a maximum a posteriori (MAP) optimization problem: it balances an action-fitting term against a prior term that keeps the seed in a high-probability region of the Gaussian latent space.
The verifier then compares the recovered seed with the owner's keyed reference.
Decoy keys put scores from different tasks and rollouts on a common scale, and aggregating the calibrated scores gives a binary ownership decision.
Ranking the same scores over candidate keys gives multi-key identification.

We evaluate our method on two policy families and two robot embodiments: $\pi_{0.5}$ and LingBot-VA on LIBERO-10, a single-arm suite, and RoboTwin, a dual-arm suite.
The no-attack evaluation answers two audit questions.
Binary ownership verification asks whether the suspect policy carries the owner's key.
Multi-key identification asks which assigned key the suspect carries when the owner has issued different keys to different releases or licensees.
We also report task success rate and episode length, so the reader can see whether the fingerprint changes robot performance.
Across the four policy--robot combinations, task success rate changes by at most a few points.
With $16$ audit rollouts, binary verification reaches TPR $1.00$ at $1\%$ FPR in all four combinations, and the same grouped evidence makes the assigned key identifiable.

We further validate the design choices and evaluate robustness.
First, we compare our seed-space mark with an output-space mark, showing why putting the signal directly on the action stream makes removal easier.
Second, we compare MAP recovery with reverse-ODE recovery (used in Tree-ring-like watermarks~\cite{wen2023treering}), showing why the verifier should fit only the observed action channels rather than guess the hidden ones.
We then test output-side post-processing attacks (clipping, smoothing, jitter, and delay) and owner-side variants (LoRA fine-tuning, pruning, and quantization).
At the canonical attack strengths, the weakest cell still reaches TPR $0.84$ at $1\%$ FPR after aggregation.
Aggressive smoothing or jitter on $\pi_{0.5}$/LIBERO-10 is the main output-side boundary, with TPR below $0.2$ even at larger rollout budgets.
Security analysis then studies false-key risk and a white-box removal attack that fine-tunes the policy against the verifier while trying to keep task success high.
Overall, the attack does not remove the fingerprint, though one setting shows a weaker identification margin.
Finally, the discussion treats behavior-cloning distillation as a boundary attack that replaces the keyed generator rather than redeploying it: the hidden fingerprint does not transfer to the student, but a less hidden variant can.

Our contributions are as follows:
\begin{itemize}[leftmargin=*]
    \item We introduce keyed latent-provenance verification, which watermarks VLA and WAM deployments without retraining the weights or adding a visible output-side action signal.
    \item We design a black-box verifier that recovers seeds from partial, post-processed executed actions and supports both ownership verification and multi-key identification.
    \item We evaluate the method across policy families and robot embodiments, including output-side attacks, owner-side variants, distillation as a process-removal boundary, and adaptive fine-tuning against the verifier.
\end{itemize}

\section{Related Work}
\label{sec:relevant-work}

\mypara{Generative robot policies}
Recent robot foundation models use generative models as action policies~\cite{black2024pi0,pi05_2025,openvla,groot_n1,gemini_robotics,lingbotva2026}.
At each planning step, the policy reads the robot observation and task command, then generates a short action chunk that drives the robot.
This paper studies the part of these systems that is common across architectures: a latent seed is sampled before generation, and the generator turns that seed into the action chunk.
The seed-to-action path may be a flow-matching solver, a diffusion sampler, or a model that first predicts an intermediate world state.

\begin{itemize}[leftmargin=*]
    \item \textit{Vision-language-action policies.}
    Vision-language-action (VLA) policies condition directly on camera images and a language instruction, then output the next chunk of robot actions~\cite{black2024pi0,pi05_2025}.
    The action generator is trained with flow matching: it starts from Gaussian noise and follows a learned vector field until it reaches an action.
    At the level used by our verifier, the vision-language backbone and the flow solver form one differentiable map from latent seed to action:
    \[
        a = F_{\theta}^{\mathrm{VLA}}(z;\, o),
    \]
    where $o$ contains the images and instruction.
    This direct noise-to-action map is the surface on which our fingerprint is injected and later recovered.

    \item \textit{World-action models.}
    World-action models use the same observation-to-action interface but route generation through a predicted future scene.
    It first predicts how the scene will change, as future video latents or frames, and then infers the action that would produce that future~\cite{lingbotva2026,ye2026world}.
    We can write the high-level structure as
    \[
        h = W_\theta(z;\, o), \qquad a = A_\theta(h;\, o),
    \]
    where $h$ is the predicted future-scene representation and $A_\theta$ maps that representation to actions.
    The path from $z$ to $a$ is deeper than in a VLA, but the composed map $A_\theta(W_\theta(z;\, o);\, o)$ is still a differentiable seed-to-action generator.
    This shared structure lets one keyed latent mechanism cover both direct VLA policies and world-action models.
\end{itemize}

\mypara{Watermarking robot and generative policies}
Robot-policy watermarking has mainly followed two routes.
Backdoor-based ownership verification trains a secret trigger into the policy so that the marked model emits a recognizable action on that trigger~\cite{guardvla}.
This approach ties the mark to the weights and requires training or fine-tuning the policy.
Output-side robot watermarking~\cite{conoco} instead targets traditional continuous-control policies by shaping exploration noise into a secret frequency band of the action stream.
This addresses a passive-sensing setting, but the keyed energy still occupies a fixed band and gives an adversary a localized filtering target.

Generative-model watermarking offers a related latent-side idea: image diffusion watermarks such as Tree-Ring, Gaussian Shading, and RingID encode ownership information in the sampling noise rather than in a visible output pattern~\cite{wen2023treering,yang2024gaussianshading,ci2024ringid}.
Our setting differs because the protected object is not an image observed in full, which means the reverse-ODE technique is no longer robust.
The verifier sees only the action channels that reach the robot after deployment, so ownership evidence must be recovered from a partial and possibly post-processed trajectory.

\section{Threat Model}
\label{sec:threat-model}

We study provenance verification for resold robot policy services.
The owner is an upstream provider who distributes a robot policy as a commercial service and later wants proof that a suspect service is reusing the protected deployment.
The adversary tries to resell or redeploy that service without being detected.
This matches commercial deployments in which a policy is exposed as a hosted API or as a licensed on-site package, while the weights and keyed sampler remain hidden~\cite{skild_brain,intelligence2025pi06vlalearnsexperience,gemini_robotics}.
In the licensed on-site case, the outside party can often wrap, forward, or post-process the service but cannot retrain the policy or replace its sealed sampler.
The outside party can invoke the policy and read the actions it executes, but cannot inspect its internals.

\mypara{Owner's goal and capability}
The owner provides the action policy and wants to protect ownership after deployment.
In the injection phase, the owner runs the policy with a keyed sampler so that the generated actions carry the fingerprint.
The owner may also release limited variants of the same service, produced by fine-tuning, pruning, or quantization.
These variants are owner-side transformations rather than adversary attacks, but verification should still succeed if a suspect service is one of them.
In the verification phase, the owner has authorized audit access to the suspect robot.
It can record only the action commands sent to the actuators, after any deployment-side post-processing.
It does not see the full model output: a model may emit $32$ dimensions, while a single-arm robot drives only $7$ and a dual-arm robot $14$.
Since non-actuated dimensions can be dropped before audit, the fingerprint must reside in the channels that reach the robot.
The owner can collect multiple rollouts from the fixed suspect service and, using its own base model, test whether those rollouts carry its key.
If the suspect is a released variant, recovery runs on the base model while the observed actions come from the variant.
We test this variant gap empirically in \Cref{sec:eval-robust}.

\mypara{Adversary's goal and capability}
The adversary obtains the owner's model only through a black-box interface, either as a hosted API or as a locally packaged service whose weights and keyed sampler remain hidden.
It redistributes that service as its own while trying to keep the robot useful and avoid the cost of training a replacement policy.
Because the adversary cannot open the weights or keyed sampler, in-scope removal attacks act on the output stream.
It may clamp action magnitudes, apply EMA smoothing, add jitter, or insert a fixed-step delay before commands reach the actuators.
Dropping non-actuated dimensions is not a separate attack because those dimensions never enter the owner's audit view.
Thus the adversary controls deployment, while the owner sees only the final action commands during audit.
We give the recovery procedure and calibration in \Cref{sec:verification}.

\mypara{Scope statement}
The in-scope suspect service still runs the protected keyed sampling process inside the black box.
Equivalently, it still invokes the owner's injection step when it generates action chunks.
Training a fresh student policy on teacher outputs and running it with a new sampler moves the adversary from redeployment to model training.
That move brings the usual data and compute burden, along with any capability loss from behavior cloning.
This is the same boundary as generation-time latent watermarks such as Tree-Ring~\cite{wen2023treering}, Gaussian Shading~\cite{yang2024gaussianshading}, and RingID~\cite{ci2024ringid}.
We map this boundary empirically in \Cref{sec:discussion-distillation,fig:distillation}, where behavior cloning exposes the trade-off between distillation resistance, Gaussian noise-law invariance, and key-space size.

\section{Method}
\label{sec:method}

The model owner wants to verify whether a suspect policy service is running the protected keyed sampling process.
We call our framework \emph{keyed latent provenance verification}.
It targets action-generative policies with continuous latent noise, including flow-matching and diffusion policies.
The method has two stages.
First, injection hides a keyed signal in the sampler's initial noise while preserving the Gaussian noise law, so the action stream does not expose an explicit mark.
Second, verification works from the audit view available after deployment: partial and possibly post-processed action commands.
From this view, the verifier estimates latent seeds, scores their alignment with candidate keyed references, and aggregates rollout evidence to decide whether the suspect policy carries the owner's key.

\mypara{Notation}
A rollout episode is indexed by $e$ and contains action chunks indexed by $c$.
At chunk $c$, the policy sees observation $o_c$ and draws latent seed noise $z_c \sim \mathcal{N}(0,I_{D_z})$.
We write the full differentiable generator from seed noise to raw action as $F_\theta$:
\[
    a_c = F_\theta(z_c;\, o_c) \in \mathbb{R}^{H \times D_{\mathrm{raw}}}.
\]
Here $D_z$ is the latent seed dimension, $H$ is the action horizon, and $D_{\mathrm{raw}}$ is the number of raw action channels emitted at each step.
The robot executes only $D_{\mathrm{env}} \le D_{\mathrm{raw}}$ of those channels.
The owner uses a secret key $k^*$ for injection.
In verification, $k$ denotes any candidate key that the owner may test.
Each episode also has a nonce $\nu_e$, which makes the keyed reference fresh across episodes.
The nonce is public rather than secret: it is part of the episode invocation context, such as an episode identifier or request identifier that the audit protocol records.
For any candidate key $k$, episode nonce $\nu_e$, and chunk $c$, a pseudorandom generator seeded by $(k,\nu_e,c)$ outputs a Gaussian reference vector $r_{k,\nu_e,c} \in \mathbb{R}^{D_z}$ in the same latent seed space as $z_c$.
When $k=k^*$, we write this reference as $r_c$ when the episode and chunk are clear.

\subsection{Fingerprint Injection}
\label{sec:injection}

The fingerprint injection procedure consists of two parts: keyed chunk selection and keyed latent perturbation.

\mypara{Keyed chunk selection}
In the injection process, not every chunk is fingerprinted.
A keyed selector $\mathcal{S}(k^*, \nu_e, c) \in \{0, 1\}$ decides which chunks carry the perturbation.
Within each episode, it marks at most $m$ chunks, with a key-dependent gap of at most $P$ chunks between selected positions.
Here, $P$ is the selector's maximum gap and $m$ is the per-episode selection cap.
Both the starting offset and the specific slots are determined by a pseudorandom generator seeded by $(k^*, \nu_e)$.
The selection schedule is therefore fixed for whoever holds the key, but unpredictable to anyone without it.

Selecting only some chunks serves two purposes.
First, it limits how often the generated action stream carries a key-correlated perturbation, which lowers the power of aggregate action-level tests and reduces task disturbance.
Second, the selection schedule is secret: an adversary without $k^*$ cannot tell which chunks carry the mark, so targeted removal would require perturbing the action stream broadly rather than editing only the marked chunks.

\mypara{Keyed latent perturbation}
At each chunk $c$, the policy sampler draws a base latent vector $z_c \sim \mathcal{N}(0, I_{D_z}) \in \mathbb{R}^{D_z}$.
If $\mathcal{S}(k^*, \nu_e, c)=1$, the chunk is selected, and the owner uses the keyed reference $r_c = r_{k^*,\nu_e,c}$ defined above.
Over fresh keyed generator outputs, the coordinates of $r_c$ are i.i.d.\ $\mathcal{N}(0, 1)$.
The selected chunk uses the fingerprinted seed:
\begin{equation}
    z_c^{\mathrm{fp}} = \sqrt{1 - \beta^2}\, z_c + \beta\, r_c,
    \label{eq:injection}
\end{equation}
where $\beta \in (0, 1]$ controls the injection strength.
Marginally over fresh base noise and keyed references, $z_c$ and $r_c$ are independent $\mathcal{N}(0, I_{D_z})$ vectors, so $z_c^{\mathrm{fp}}$ is again $\mathcal{N}(0, I_{D_z})$: mean $0$ and covariance $(1-\beta^2)I_{D_z} + \beta^2 I_{D_z} = I_{D_z}$.
The injection therefore leaves the sampler's latent-noise distribution unchanged: without $k^*$, no statistical test on individual sampler draws can separate fingerprinted from base noise.
If $\mathcal{S}(k^*, \nu_e, c)=0$, the sampler uses the unmodified seed $z_c$.
The policy then passes the chosen seed through flow matching or diffusion to generate the action chunk.

This placement is why the fingerprint is injected into the sampler noise rather than added to the executed action.
An output-space mark is easy for the owner to define and score, but it remains exposed in the action stream.
If that mark concentrates energy in a narrow frequency band, an adversary can locate the band and remove it directly.
Latent injection instead places the reference before the model's nonlinear generator.
The policy mixes the keyed perturbation while generating the action, so the executed action does not expose the reference as a simple additive signal and the fingerprint energy spreads across the action spectrum (\Cref{fig:spectrum}).
\Cref{sec:eval-design-injection} tests this difference with an output-space baseline.

\subsection{Fingerprint Verification}
\label{sec:verification}

In the verification stage, the verifier turns observed actions into evidence about the key.
For a candidate key $k$, it estimates the latent seed that entered the generator behind each observed action chunk, scores that estimate against the references generated from $k$, and aggregates these scores across rollouts.
For ownership detection, the candidate key of interest is $k^*$.
Below, we describe each step.

\mypara{Observation model}
The verifier cannot access the model's internal noise directly.
It records only the actions the robot executes.
These actions differ from the policy's raw output in two ways.
First, the environment runs only $D_{\mathrm{env}}$ out of the $D_{\mathrm{raw}}$ channels, for example, for a single-arm robot suite, $D_{\mathrm{env}} = 7$ while $D_{\mathrm{raw}} = 32$.
Second, the deployment may post-process the executed stream before it reaches the actuators.
In the experiments, this includes clipping, exponential smoothing, additive jitter, and fixed-step delay, but the observation model treats these as instances of a general post-processing map.
Let $\tilde z_c$ be the latent seed that actually enters the generator at chunk $c$: it is $z_c^{\mathrm{fp}}$ for a selected chunk and the unmodified base seed otherwise.
The verifier therefore observes:
\begin{equation}
    y_c = g\!\left(P_C(F_\theta(\tilde z_c;\, o_c))\right) + \epsilon_c,
    \label{eq:obs-model}
\end{equation}
where $P_C$ projects the raw action chunk from $\mathbb{R}^{H \times D_{\mathrm{raw}}}$ to the environment-visible channels in $\mathbb{R}^{H \times D_{\mathrm{env}}}$, $g(\cdot)$ models deployment-side post-processing (clipping, smoothing, jitter, delay), and $\epsilon_c$ captures measurement noise at chunk $c$.
In the full-observation setting, $P_C = I$ and $g$ is the identity function.

\mypara{MAP latent recovery}
Recovering $\tilde z_c$ from this partial and possibly post-processed observation is the main challenge in the verification stage.
The obvious baseline is to run the generative ODE backward from the observed action.
This is ill-posed when $y_c$ is partial: the $D_{\mathrm{env}}$ observed channels leave the remaining $D_{\mathrm{raw}} - D_{\mathrm{env}}$ raw channels underdetermined.
A reverse-ODE method (like Tree-ring~\cite{wen2023treering}) must first guess those missing channels before running backward, and errors in that completion can move the reverse trajectory away from the model's training convention.
We instead pose latent recovery as a MAP estimation problem: find the noise $\hat{z}_c$ that best fits the observed $D_{\mathrm{env}}$ channels while staying close to the Gaussian prior:
\begin{equation}
    \hat{z}_c = \arg\min_z \;
    \underbrace{\frac{1}{2\sigma_{\mathrm{obs}}^2} \left\| g\!\left(P_C(F_\theta(z;\, o_c))\right) - y_c \right\|_2^2}_{\text{observation likelihood}}
    + \underbrace{\frac{\lambda_z}{2} \|z\|_2^2}_{\text{Gaussian prior}}.
    \label{eq:map}
\end{equation}
Here $\sigma_{\mathrm{obs}}$ sets the observation-noise scale, and $\lambda_z$ controls how strongly the Gaussian prior penalizes large seeds.
The Gaussian prior keeps the recovered seed in a high-density part of the latent space, so the optimizer cannot explain missing action channels by drifting to unlikely latent values.
Unlike reverse ODE inversion, MAP recovery does not require the verifier to guess whether the training pipeline padded, masked, or otherwise filled channels that the robot never executed.
We estimate $\tilde z_c$ by starting gradient descent from several random initial seeds.
Each run sends its current seed through $F_\theta$, compares the predicted executed channels with $y_c$, and updates the seed to reduce the loss in \Cref{eq:map}.
We keep the run with the lowest final loss.
Implementation choices for $\beta$, $P$, $m$, $\lambda_z$, and $\sigma_{\mathrm{obs}}$ are given in \Cref{app:method-hyperparameters}.
\Cref{sec:eval-design-recovery} tests this choice against reverse ODE inversion.

\mypara{Key-conditioned scoring}
For each episode $e$ and candidate key $k$, the verifier applies the same selector $\mathcal{S}(k,\nu_e,c)$ to choose the chunks that this key would have marked.
Let $M_e(k)$ be the number of selected chunks in that episode.
The verifier orders those chunks by time and writes the recovered latent seeds as $\hat{z}_{e,1:M_e(k)}$ and the matching references as $r_{k,1:M_e(k)}$.
This shorthand uses the same nonce $\nu_e$ and the same chunk positions on both sides.
For the lag notation below, we write these ordered sequences as $\hat z_e(k)$ and $r_{k,e}$.
It then computes a score:
\begin{equation}
    s_e(k) = \psi\!\left(\hat{z}_{e,1:M_e(k)},\; r_{k,1:M_e(k)}\right),
\end{equation}
where $\psi(\hat z, r) = r^\top \hat z$ is the matched filter.
The inner product is taken after flattening the selected seed vectors, and it measures how strongly the recovered seeds align with the keyed reference.
In the ideal additive-Gaussian model, this matched filter is the Neyman--Pearson optimal score~\cite{kay1998detection}.
When the controller introduces a timing delay, the verifier replaces $\psi$ with the lag-conditioned score defined in the synchronization search below.

\mypara{Z-score construction}
The raw scores $s_e(k)$ vary across episodes because tasks and trajectory lengths differ.
To put scores from different episodes on the same scale, the verifier also scores a bank of $J$ decoy keys.
For a candidate key $k$, choose decoy keys $\{k_j\}_{j=1}^{J}$ that are distinct from $k$ and sampled independently from the same key space.
The verifier computes $s_e(k_j)$ for every decoy key in the same episode, using that decoy key's selected chunks, then uses these decoy scores to estimate the episode's wrong-key mean and scale:
\begin{equation}
    \begin{aligned}
    \mu_e^{-}(k)
    &= \frac{1}{J}\sum_{j=1}^{J}s_e(k_j), \\
    \sigma_e^{-}(k)
    &=
    \Bigg[
    \frac{1}{J-1}\sum_{j=1}^{J}
    \left(s_e(k_j)-\mu_e^{-}(k)\right)^2
    \Bigg]^{1/2}.
    \end{aligned}
    \label{eq:decoy-key-stats}
\end{equation}
The calibrated score is:
\begin{equation}
    Z_e(k) = \frac{s_e(k) - \mu_e^{-}(k)}{\sigma_e^{-}(k) + \epsilon_Z}.
    \label{eq:zscore}
\end{equation}
The small constant $\epsilon_Z$ only prevents division by zero.
This normalization has a useful property: because the decoy references are generated by the same procedure as the owner's reference and differ only in the key value, $\mu_e^{-}(k)$ and $\sigma_e^{-}(k)$ measure what recovery and matching return for the \emph{wrong} key in that episode.
As a result, $Z_e(k^*)$ reflects only the response specific to the owner's key, not any correlation that would appear for any key due to the rollout's structure.
Here $H_0$ means that the suspect policy does not carry candidate key $k$, and $H_1$ means that it carries and preserves that key.
Under $H_0$, the decoy normalization makes $Z_e(k)$ approximately centered with unit scale; under $H_1$, the carried key produces a positive shift.

\mypara{Multi-trajectory evidence aggregation}
Because of task variability, controller effects, partial observability, and stochastic execution, a single rollout gives weak or ambiguous evidence about the key.
Summing the calibrated scores over a group $G$ of rollouts amplifies the true-key signal while averaging out per-episode noise:
\begin{equation}
    T_G(k) = \sum_{e \in G} Z_e(k).
    \label{eq:aggregation}
\end{equation}

$G$ can contain same-task or cross-task rollouts; the group size $|G|$ controls how many rollouts the verifier collects from the suspect.
Under $H_0$, $T_G(k)$ stays near zero.
Under $H_1$ for the owner's key, each episode contributes a positive increment, so $T_G(k^*)$ grows with $|G|$.
The final decision compares $T_G(k^*)$ against a threshold set at the $(1-\alpha_{\mathrm{FPR}})$ quantile of the decoy-key distribution of $T_G$, targeting a false positive rate of $\alpha_{\mathrm{FPR}}$.

\mypara{Synchronization search}
The verifier includes a small alignment step before scoring.
For a candidate lag $\ell$, let $\psi_\ell(\hat z_e(k), r_{k,e})$ be the matched-filter score after shifting the recovered seed representation against the keyed reference and keeping only coordinates with valid control-time overlap; $\psi_0$ is the lag-zero score above.
The verifier picks the lag $\ell^*$ that maximizes the keyed response summed over $G$:
\begin{equation}
    \ell^* = \arg\max_{0 \le \ell \le L}\; \sum_{e \in G} \psi_\ell\!\left(\hat z_e(k^*),\; r_{k^*,e}\right),
    \label{eq:lagsearch}
\end{equation}
where $L$ is the maximum lag searched.
Every episode is then scored at this single lag: $s_e(k) = \psi_{\ell^*}(\hat z_e(k), r_{k,e})$.
Since the same $\ell^*$ is applied to the true key and all $J$ decoys, the decoy-key distribution uses the same alignment choice as the owner key.
We set $L = \max(\lceil 0.1\,H\rceil, 4)$, where $H$ is the chunk horizon; this keeps the searched shifts small relative to the action window while still covering short multi-step offsets.

\mypara{Detection Rate}
After MAP recovery, decoy-key calibration, and optional lag alignment, each episode contributes one calibrated score $Z_e(k)$.
The group statistic $T_G(k^*)$ then follows a simple scaling law: as $|G|$ grows, detection depends only on the per-episode signal-to-noise of these calibrated scores.

\begin{proposition}[Detection rate of group aggregation]
\label{prop:rate}
Assume episodes in $G$ are independent.
Under $H_1$, let the per-episode calibrated score have mean $\mu$ and variance $\sigma_1^2$.
Under $H_0$, let it have mean $0$ and variance $\sigma_0^2$.
Then the d-prime of $T_G(k^*)=\sum_{e\in G}Z_e(k^*)$ is
\[
d'(|G|) = \sqrt{|G|}\cdot\frac{\mu}{\sqrt{(\sigma_0^2+\sigma_1^2)/2}},
\]
and under a Gaussian approximation the TPR at FPR $\alpha_{\mathrm{FPR}}$ is $\Phi\!\big(d'(|G|) - \Phi^{-1}(1-\alpha_{\mathrm{FPR}})\big)$.
\end{proposition}

The $\sqrt{|G|}$ scaling shows when aggregation can rescue a weak per-episode signal.
If $\mu$ is small but positive, $d'$ still grows without bound, so a moderate $|G|$ reaches near-perfect detection.
A fingerprint-free model ($\mu=0$) gains nothing from aggregation.
Inverting the relation gives the rollout budget needed for a target $(\mathrm{TPR},\mathrm{FPR})$:
\[
|G| \ge
\left(
\frac{
\left[\Phi^{-1}(\mathrm{TPR})+\Phi^{-1}(1-\mathrm{FPR})\right]
\sqrt{(\sigma_0^2+\sigma_1^2)/2}
}{\mu}
\right)^2 .
\]
We compare this prediction against the bootstrap measurements in \Cref{sec:eval-verification}.

\section{Verification Evaluation}
\label{sec:eval}

\subsection{Experimental Setup}
\label{sec:setup}
\begin{figure}[t]
\centering
\begin{minipage}{0.95\columnwidth}
\centering
\includegraphics[width=\linewidth]{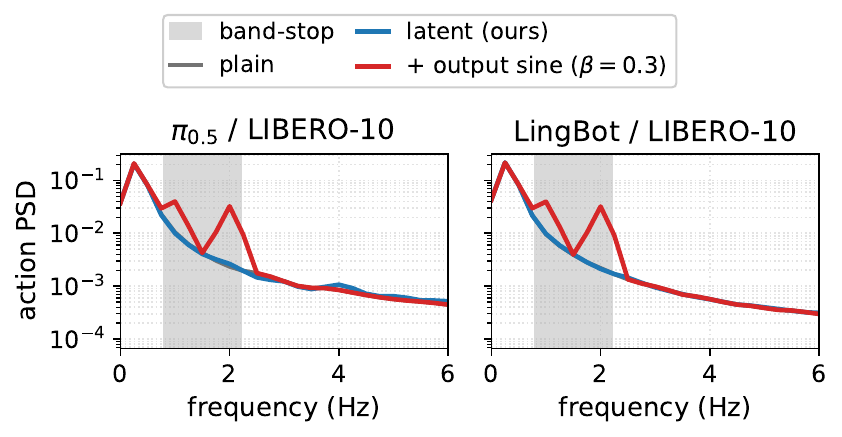}\\[-0.35em]
\textbf{(a)} Executed-action spectrum.
\end{minipage}
\vspace{0.45em}
\begin{minipage}{\columnwidth}
\centering
\includegraphics[width=\linewidth]{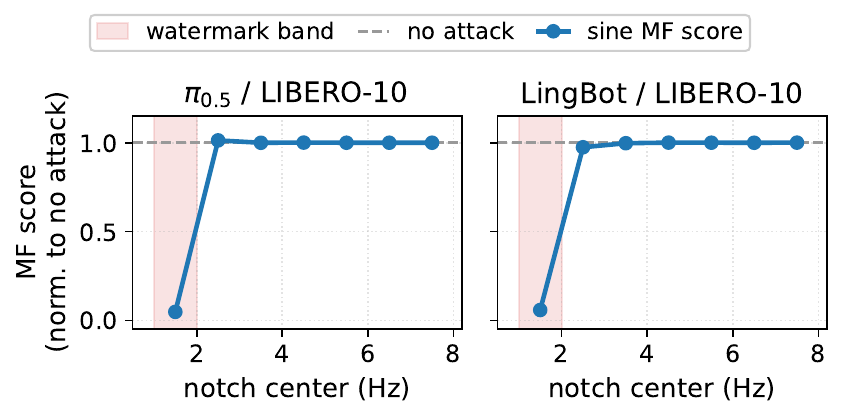}\\[-0.35em]
\textbf{(b)} Band-stop removal sweep.
\end{minipage}
\caption{Output-level baseline. Panel (a) plots the executed-action spectrum; the shaded region is the $1$--$2$\,Hz sine-watermark band. Panel (b) moves a fixed-width band-stop filter across frequencies and reports the matched-filter score after filtering.}
\label{fig:output-level-baseline}
\label{fig:spectrum}
\label{fig:bandstop-sweep}
\end{figure}
\begin{table*}[t]\centering
\caption{Overall verification under partial observation with MAP recovery. Utility columns give task success rate (SR) and mean episode steps for plain vs.\ fingerprinted rollouts. AUC is per-episode ($|G|{=}1$); AUC$_{16}$ and TPR at FPR$=1\%$ use group budget $|G|{=}16$. Detection columns report the no-attack condition and the average over canonical-strength removal attacks. 
}
\label{tab:main}\small
\begin{tabular}{ll cc c ccc ccc}
\toprule
& & \multicolumn{3}{c}{Utility (clean)} & \multicolumn{3}{c}{Clean} & \multicolumn{3}{c}{Robust (atk.\ avg)} \\
\cmidrule(lr){3-5}\cmidrule(lr){6-8}\cmidrule(lr){9-11}
Model & Dataset & SR$_{\mathrm{pl}{\to}\mathrm{fp}}$ & $\Delta$SR & Steps$_{\mathrm{pl}{\to}\mathrm{fp}}$ & AUC & AUC$_{16}$ & TPR@1\% & AUC & AUC$_{16}$ & TPR@1\% \\
\midrule
LingBot & LIBERO-10 & 0.94$\to$0.96 & +0.02 & 277$\to$271 & 1.000 & 1.000 & 1.000 & 1.000 & 1.000 & 1.000 \\
LingBot & RoboTwin & 0.57$\to$0.53 & -0.04 & 184$\to$187 & 1.000 & 1.000 & 1.000 & 0.999 & 1.000 & 1.000 \\
pi0.5 & LIBERO-10 & 0.96$\to$0.96 & +0.00 & 261$\to$266 & 0.901 & 1.000 & 1.000 & 0.798 & 0.991 & 0.840 \\
pi0.5 & RoboTwin & 0.50$\to$0.57 & +0.07 & 412$\to$337 & 0.852 & 1.000 & 1.000 & 0.854 & 1.000 & 1.000 \\
\bottomrule
\end{tabular}
\end{table*}

\mypara{Evaluation design}
This evaluation asks whether the same verifier works across two sources of variation: the action generator and the robot embodiment.
We therefore test one flow-matching VLA, $\pi_{0.5}$, and one world-action model, LingBot-VA, on LIBERO-10 as the single-arm suite and RoboTwin as the dual-arm suite.
The model determines the raw action dimension, and the robot suite determines which of those channels are executed.
The executed/raw channel counts are $7/32$ for $\pi_{0.5}$/LIBERO-10, $7/30$ for LingBot-VA/LIBERO-10, $14/32$ for $\pi_{0.5}$/RoboTwin, and $16/30$ for LingBot-VA/RoboTwin.
Together these four model--suite cells let the results be read against both generator family and partial observation.

\mypara{Common protocol}
Each cell has a plain rollout pool and a fingerprinted rollout pool.
The plain pool runs the same policy without latent injection.
It supplies the utility comparison and the no-injection false-positive check.
The fingerprinted pool runs with the owner's key $k^*$: the selector chooses up to five action chunks per episode, and each selected chunk receives a reference with strength $\beta=1.0$ and four Gaussian tones in the $1$--$2$\,Hz band.
Unless stated otherwise, the verifier observes only the executed channels, recovers one latent estimate for each selected chunk with MAP, and scores the recovered latents against the owner's key.
It also scores the same episode against $J=32$ decoy keys, which gives the wrong-key mean and scale used to form the calibrated episode score $Z_e(k^*)$.

The query budget $|G|$ is the number of suspect episodes in one provenance decision.
To report a budget, we repeatedly form groups of $|G|$ fingerprinted episodes and sum their owner-key calibrated scores as $T_G(k^*)$.
For TPR at FPR$=1\%$, the threshold is set from grouped decoy-key scores, and the true-positive rate is measured on these grouped owner-key scores.
The confidence bands in the figures come from bootstrapping the episode-level score pools before forming these groups.

\subsection{Injection-Site Validation}
\label{sec:eval-design-injection}
\begin{figure*}[t]
\centering
\includegraphics[width=0.95\linewidth]{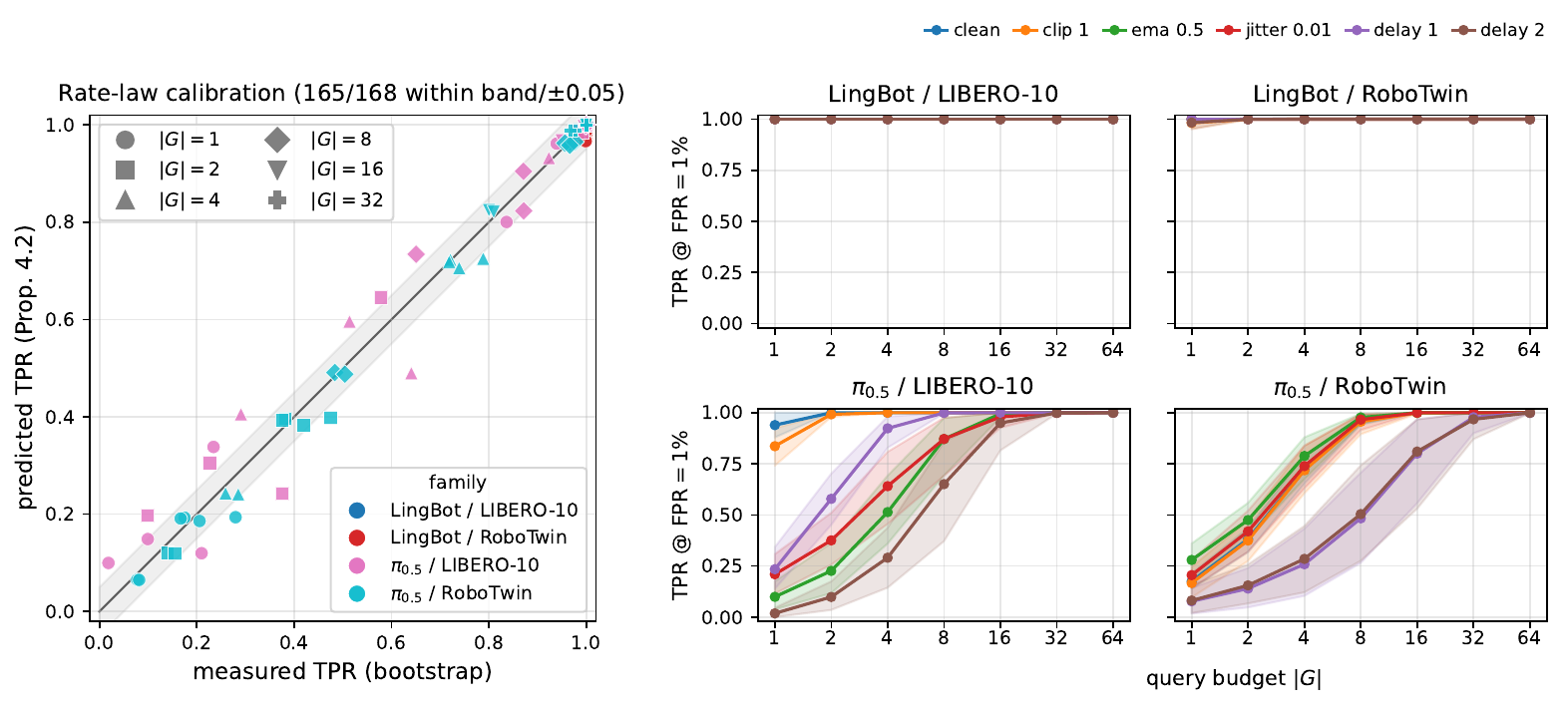}
\caption{Detection power and rate-law calibration under partial observation with MAP recovery. \textbf{Right:} true positive rate vs.\ query budget $|G|$ at FPR$=1\%$, with one panel per model--suite cell and one line per attack condition. Bands are 5--95\% bootstrap intervals. \textbf{Left:} predicted TPR from \Cref{prop:rate} vs.\ measured TPR for each model--suite cell, attack condition, and query budget. The diagonal is $y{=}x$, the shaded envelope is $\pm0.05$ TPR, colors identify model--suite cells, and markers identify $|G|$.}

\label{fig:tpr-vs-g}
\end{figure*}

\begin{figure}[h]
\centering
\includegraphics[width=\columnwidth]{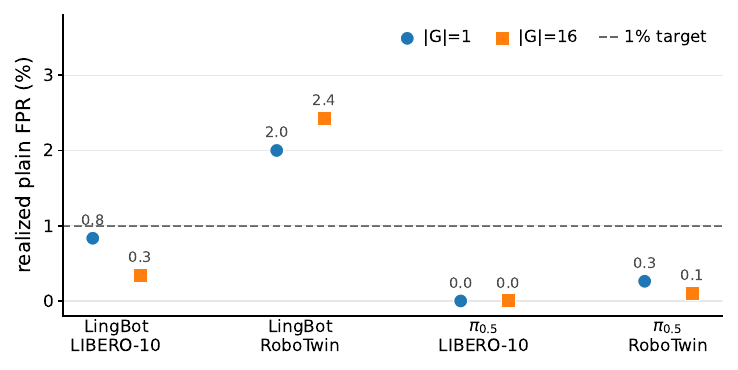}
\caption{Decoy-key threshold calibration at the nominal $1\%$ operating point. Each point is the realized false-positive rate on plain rollouts without injection when the threshold is set from decoy-key scores, shown for $|G|{=}1$ and $|G|{=}16$. The dashed line is the nominal target; points below it are conservative, while points above it are anti-conservative.}
\label{fig:neg-control-h0}
\end{figure}

This validation tests why the fingerprint is injected in latent space rather than directly into the executed action.
The output-level alternative adds the keyed reference to the action that reaches the robot, then verifies it with the same matched-filter machinery.
We make this baseline strong rather than artificial: the verifier knows the keyed sine family, scores each episode against the true key and $32$ decoy keys, and uses the same LIBERO-10 action streams used elsewhere in the paper ($100$ $\pi_{0.5}$ episodes and $200$ LingBot-VA episodes).
The sine occupies the $1$--$2$\,Hz band at strength $\beta{=}0.05$.
The removal attack is a 4th-order Butterworth band-stop filter.
The default notch, $0.8$--$2.2$\,Hz, covers the watermark band, and a fixed-width sweep moves the notch to test whether removal is localized rather than a generic effect of filtering.
\begin{table}[b]\centering
\caption{Latent recovery on $\pi_{0.5}$/LIBERO-10: MAP vs.\ reverse-ODE, by
observation regime and, for partial-observation ODE, by pad-completion assumption.
Columns are per-chunk separation AUC, Cohen's $d$, and recovered-noise norm
$\|\hat z\|_2$.}
\label{tab:recovery}\small
\resizebox{\linewidth}{!}{\begin{tabular}{ll ccc}
\toprule
Obs. & Recovery & AUC & Cohen $d$ & $\|\hat z\|_2$ \\
\midrule
\multirow{2}{*}{full}
 & ODE                              & $0.809$          & $1.26$          & --- \\
 & MAP                              & $\mathbf{0.994}$ & $\mathbf{3.34}$ & --- \\
\midrule
\multirow{5}{*}{partial}
 & ODE, no pad (underdet.)          & $0.492$          & $-0.03$         & --- \\
 & ODE, pad $V{=}0$ (matched)       & $\mathbf{0.969}$ & $\mathbf{2.30}$ & $4.1$ \\
 & ODE, pad $V{=}1$ (mismatch)      & $0.678$          & $0.62$          & $79$ \\
 & ODE, secret random pad           & $0.508$          & $0.02$          & $79$ \\
 & MAP (pad-agnostic)               & $0.775$          & $1.07$          & --- \\
\bottomrule
\end{tabular}
}\end{table}

\begin{figure*}[t]
\centering
\includegraphics[width=0.95\textwidth]{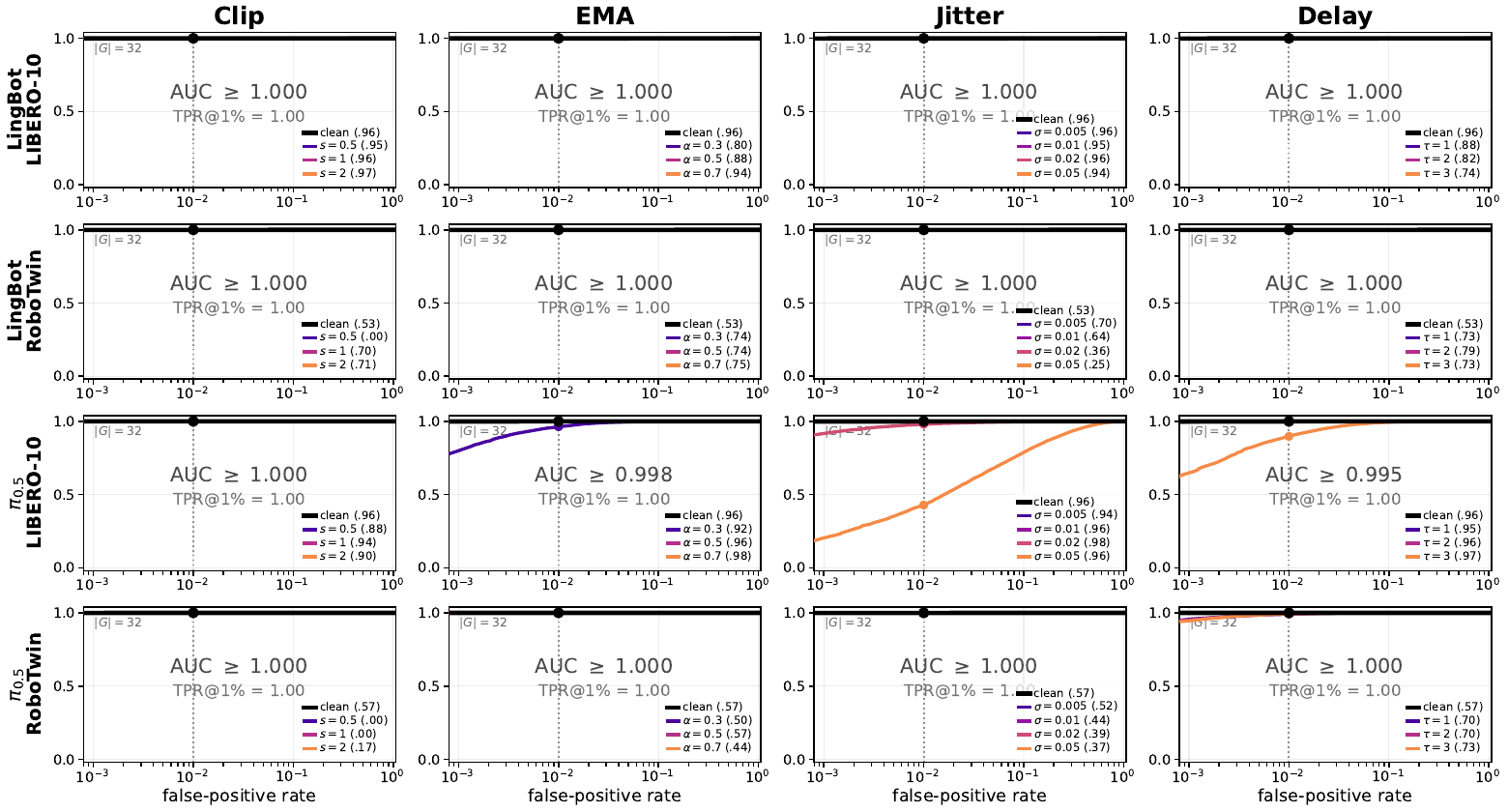}
\caption{ROC under output-side removal attacks at group budget $|G|{=}32$. Rows are model--suite cells; columns are attack families; black is the no-attack condition and purple$\to$orange means weaker$\to$stronger attack. The dotted vertical marks the $1\%$ FPR target and the dot marks the operating point on each curve; parentheses give the fingerprinted policy's success rate. Where all curves in a panel are near-perfect, the panel text gives the minimum AUC and TPR@1\% instead of overlapping flat curves.}
\label{fig:attack-combined}
\end{figure*}

\Cref{fig:output-level-baseline} shows the main failure mode of output-level marking: the mark is easy for the owner to score because it is also easy for an adversary to locate.
The output mark creates visible spectral peaks in the shaded $1$--$2$\,Hz band.
When the notch misses that region, the matched-filter score remains high.
When the notch covers it, the score collapses.
At the provenance operating point, the default notch cuts the sine score by about $90\%$.
Single-shot TPR@$1\%$ drops from $0.82$ to $0.17$ on $\pi_{0.5}$ and from $0.95$ to $0.18$ on LingBot-VA.
The action edit is modest, about $0.06$ per-coordinate RMSE, and is far below the perturbation strengths that damage task success in \Cref{fig:attack-combined}.

This experiment isolates the weakness that latent injection is meant to avoid.
An output-space mark remains an exposed action artifact.
It is easy to score, but the same visibility gives the adversary a cheap removal handle.
By injecting the keyed reference before the generator, our method lets the policy mix the signal before it reaches the robot.
The executed action therefore does not present one narrow band that can be removed without otherwise changing the policy behavior.

\subsection{Recovery Validation}
\label{sec:eval-design-recovery}

This validation tests why the verifier uses MAP rather than reverse ODE inversion.
Reverse ODE can run backward from a full raw action chunk, but the audit view is usually partial.
On $\pi_{0.5}$/LIBERO-10, for example, the verifier sees $7$ of $32$ raw channels, so reverse ODE must either invert from an underdetermined endpoint or guess the $25$ hidden channels.
MAP avoids that completion step by fitting only the observed channels while a Gaussian prior keeps the recovered seed plausible.

\Cref{tab:recovery} compares the two recovery rules on the same $50$ fingerprinted and $50$ plain $\pi_{0.5}$/LIBERO-10 rollouts from the clean base policy ($g{=}\mathrm{id}$).
Only the observation regime and recovery rule change.
AUC and Cohen's $d$ measure per-chunk separation between recovered fingerprinted latents and the plain pool.
For partial-observation ODE, we vary how the missing channels are filled before inversion.
The no-pad row leaves the endpoint underdetermined.
The matched row uses the training pad $V{=}0$, the mismatch row uses the wrong constant pad $V{=}1$, and the secret-random row fills the channels with a random value unknown to the verifier.

The results show that reverse ODE works only when hidden channels are completed correctly.
With full observation, both methods recover the fingerprint, and MAP separates more cleanly (AUC $0.994$ vs.\ $0.809$).
With partial observation, ODE falls to chance without a completion rule (AUC $0.492$) and becomes strong only under the oracle matched pad (AUC $0.969$).
Wrong or secret pads break this advantage: AUC drops to $0.678$ with $V{=}1$ and $0.508$ with a secret random pad.
MAP is not the best oracle row, but it is pad-agnostic, still produces a keyed signal (AUC $0.775$), and aggregates to TPR@1\%${=}1.0$ at $|G|{=}16$ in the main verifier (\Cref{fig:tpr-vs-g}).

\subsection{Verification Results}
\label{sec:eval-verification}

The main clean-policy result is that latent injection leaves task utility nearly unchanged and makes the owner key detectable in all four model--suite cells.
\Cref{tab:main} shows that task success rate changes by at most a few points after injection, and mean episode length stays comparable between plain and fingerprinted rollouts.
Detection is already strong from one episode and becomes reliable once the verifier aggregates rollouts.
The LingBot-VA cells separate the true key almost perfectly from one episode (AUC ${\geq}0.999$ on both LIBERO-10 and RoboTwin).
For $\pi_{0.5}$, single episodes are weaker but still above chance: AUC $0.90$ on LIBERO-10 and $0.85$ on RoboTwin.
Aggregation closes this gap: in \Cref{tab:main}, all four cells reach AUC$_{16}{=}1.0$ and TPR@1\%${=}1.0$ in the no-attack condition.
\Cref{fig:tpr-vs-g} shows the same trend as the query budget grows.

\mypara{Comparison with the predicted rate}
The rate law predicts how many rollouts are needed once the per-episode score distribution is known.
In \Cref{fig:tpr-vs-g}, the measured TPR matches the Gaussian prediction of \Cref{prop:rate} within the 5--95\% bootstrap band, or within $0.05$ TPR, in 165 of 168 model--suite, attack, and $|G|$ settings.
The remaining misses occur at low budgets in the $\pi_{0.5}$/LIBERO-10 cell, where the verifier sees only seven executed channels: \texttt{jitter} $0.01$ at $|G|{=}2$, and \texttt{delay} $\tau{=}2$ at $|G|{\in}\{1,2\}$.
These are small-budget deviations where the symmetric-Gaussian approximation is most sensitive to per-episode skew, and the gap disappears as aggregation pulls $T_G$ toward normality.
This agreement supports using \Cref{prop:rate} to plan the query budget.

\mypara{Decoy-key calibration vs.\ literal $H_0$}
The verifier sets its threshold from decoy keys, but the literal $H_0$ is a no-injection suspect policy that does not carry $k^*$.
We therefore check whether a threshold calibrated to nominal FPR$=1\%$ on decoy keys also controls false positives on plain rollouts.
It does for three of the four model--suite cells: the realized plain-FPR stays at or below the nominal value in \Cref{fig:neg-control-h0}.
For these cells, the decoy-key null is a conservative stand-in for the literal $H_0$, so the reported detection rates are lower bounds.

The exception is LingBot-VA/RoboTwin, where the realized plain-FPR reaches $2.0\%$ at $|G|{=}1$ and $2.4\%$ at $|G|{=}16$ at the nominal $1\%$ point.
Its plain rollouts carry chunk-level correlation that aligns weakly with the keyed reference even without injection, which the decoy-key threshold cannot fully separate.
This makes the fixed-threshold TPR mildly anti-conservative on that cell.
The same effect can also inflate realized FPR under output post-processing, especially jitter, so when this drift matters we read the threshold-free AUC in \Cref{tab:main,tab:robust-detection}.

\subsection{Robustness to Attacks and Owner Variants}
\label{sec:eval-robust}

\mypara{Output-level attack setup}
Output-side attacks test the adversary class that can alter executed actions but cannot open the keyed sampler.
We use four attack families from \Cref{sec:threat-model}: clipping, exponential smoothing, additive jitter, and fixed-step delay.
Each attack is applied to both fingerprinted and plain rollouts, so the AUC measures the attack's effect on detecting the key rather than a generic shift in task distribution.
For clipping, smoothing, and jitter, we sweep strengths where rollouts are available.
\texttt{delay} is collected for all four model--suite cells.
All cells use the same verifier: decoy-key calibration, group aggregation, and a $1\%$ FPR operating point.

\mypara{Output-level: results}
The main pattern is that output post-processing usually fails to remove the fingerprint unless it also destroys a large part of the action signal.
On both LingBot-VA cells, every output attack stays at the detection ceiling: per-episode AUC is at least $0.99$, and TPR@1\% is $1.00$ at $|G|{=}16$.
On the $\pi_{0.5}$ cells, the attacks split into two regimes.
If the attack also degrades the policy, as clipping does on $\pi_{0.5}$/RoboTwin, suppressing the fingerprint costs the adversary the capability it is trying to keep.
The harder case is $\pi_{0.5}$/LIBERO-10, where smoothing, jitter, and delay can perturb the action stream while keeping task success rate near $0.95$.

Even in this harder cell, the failure modes are specific rather than general.
A delay mostly shifts the signal in time, so synchronization search realigns it.
With the search enabled, a one-step delay rises from TPR $0.28$ to $1.00$ at $|G|{=}16$, and two- or three-step delays recover once the group budget reaches $|G|{=}32$ or $64$ (\Cref{tab:lagsearch}).
Smoothing is different because it attenuates the signal itself.
The default smoothing attack gives the lowest robust-average AUC in \Cref{tab:main} ($0.80$), but its per-episode signal remains above chance and reaches the ceiling by $|G|{=}64$.
Only the strongest smoothing and jitter settings both preserve task success and cut per-episode AUC to about $0.55$; for these two settings, TPR at $1\%$ FPR stays below $0.2$ even at $|G|{=}64$ (\Cref{tab:robust-detection}).
These two settings mark the boundary of output-level robustness.

\begin{table}[t]\centering
\caption{Owner-side variant robustness. AUC$_1$ is per episode; AUC$_{16}$ and TPR use $|G|{=}16$ at FPR$=1\%$. }
\label{tab:weight-level}\small
\setlength{\tabcolsep}{2.5pt}
\begin{tabular}{@{}ll l ccc@{}}
\toprule
Model & Dataset & Edit & AUC$_1$ & AUC$_{16}$ & TPR \\
\midrule
LingBot & LIBERO-10 & LoRA & 1.00 & 1.00 & 1.00 \\
LingBot & LIBERO-10 & \texttt{prune30} & 1.00 & 1.00 & 1.00 \\
LingBot & LIBERO-10 & \texttt{int8} & 1.00 & 1.00 & 1.00 \\
\cmidrule(lr){1-6}
LingBot & RoboTwin & LoRA & 0.83 & 1.00 & 1.00 \\
LingBot & RoboTwin & \texttt{prune30} & 1.00 & 1.00 & 1.00 \\
LingBot & RoboTwin & \texttt{int8} & 1.00 & 1.00 & 1.00 \\
\cmidrule(lr){1-6}
$\pi_{0.5}$ & LIBERO-10 & LoRA & 1.00 & 1.00 & 1.00 \\
$\pi_{0.5}$ & LIBERO-10 & \texttt{prune30} & 0.99 & 1.00 & 1.00 \\
$\pi_{0.5}$ & LIBERO-10 & \texttt{int8} & 1.00 & 1.00 & 1.00 \\
\cmidrule(lr){1-6}
$\pi_{0.5}$ & RoboTwin & LoRA & 0.87 & 1.00 & 1.00 \\
$\pi_{0.5}$ & RoboTwin & \texttt{prune30} & 0.68 & 0.97 & 0.56 \\
$\pi_{0.5}$ & RoboTwin & \texttt{int8} & 0.76 & 1.00 & 0.94 \\
\bottomrule
\end{tabular}
\end{table}

\mypara{Owner-side variant setup}
Weight edits test a variant gap rather than an adversary removal attack: the verifier uses the owner's base model, but the observed actions come from a released descendant.
We fine-tune fingerprinted descendants with LoRA adapters on additional task data and evaluate them with the same verifier as the base policy.
The comparison asks whether the keyed latent perturbation still affects the descendant's actions strongly enough for MAP recovery to find it.
\Cref{tab:weight-level} lists the owner-side variant rows in the main text; the full swept output-attack rows remain in \Cref{tab:robust-detection}.

\mypara{Fine-tuning: results}
The fingerprint survives LoRA fine-tuning in every cell.
At $|G|{=}16$, all fine-tuned descendants reach group AUC $1.000$ and TPR@1\%${=}1.00$.
The cells differ only before aggregation: the weakest single-episode cell is LingBot-VA/RoboTwin, with AUC$_1{=}0.83$.
LingBot-VA/RoboTwin is the hardest fine-tuning cell because its single-task objective and deeper diffusion backbone reshape more of the path from latent seed to action.
Aggregation removes that ambiguity, so fine-tuning does not prevent detection at the standard query budget.

\mypara{Compression: setup}
Compression tests whether post-training size reductions break the path from keyed latent seed to action.
We evaluate magnitude pruning and int8 quantization on the fingerprinted policy.
For LingBot-VA, the edit touches the diffusion transformer.
For $\pi_{0.5}$, we also report an action-only compression that edits the action expert and projection heads while leaving the vision-language backbone intact.

\mypara{Compression: results}
Compression leaves the fingerprint recoverable except when the edit also damages the policy backbone.
On every LIBERO-10 cell and both LingBot-VA cells, group AUC stays at $1.000$ and per-episode AUC stays at least $0.94$.
The only sensitive case is $\pi_{0.5}$/RoboTwin, where compressing the whole model also compresses the vision-language backbone.
That edit halves task success (SR${=}0.41$) and drops per-episode AUC to $0.60$, but the fingerprint is still recoverable after aggregation ($\mathrm{AUC}_{16}{=}0.88$ pruned, $0.98$ quantized).
When the compression is restricted to the action-side modules, the realistic case summarized in \Cref{tab:weight-level}, the policy stays usable and the fingerprint remains separable after aggregation ($\mathrm{AUC}_{16}{=}0.97/1.00$).
The loss therefore comes from damaging the backbone, not from erasing the fingerprint.

\subsection{Ablations}
\label{sec:ablation}
We leave two verifier ablations to the appendix.
\Cref{app:lagsearch} isolates the synchronization search on the \texttt{delay} attack: on the hardest delayed cell, a one-step delay recovers from TPR $0.28$ to $1.00$ at $|G|{=}16$ after alignment.
\Cref{app:aggregation-mode} compares same-task and cross-task aggregation; on the LingBot-VA cells, where both comparisons are meaningful, the curves nearly overlap.
Thus, alignment helps when actions are shifted in time, while aggregation does not rely on mixing tasks.

\section{Identification Evaluation}
\label{sec:eval-identification}

The verification experiment asks a binary question: does the suspect policy carry the owner's key $k^*$?
Identification asks the attribution question that arises when the owner assigns different keys to different licensees or releases: which key does the suspect carry?
The verifier does not need new rollouts for this task.
For every candidate key $k$, it already computes an episode score $Z_e(k)$ and sums those scores over a rollout group as $T_G(k)$.
The identification gallery in our experiments contains $33$ candidate keys: the true key $k^*$ and the same $J{=}32$ decoy keys used for calibration.
For decoy keys, we calibrate leave-one-out against the other decoys so all gallery scores live on one scale.
\begin{figure*}[t]
\centering
\includegraphics[width=0.92\textwidth]{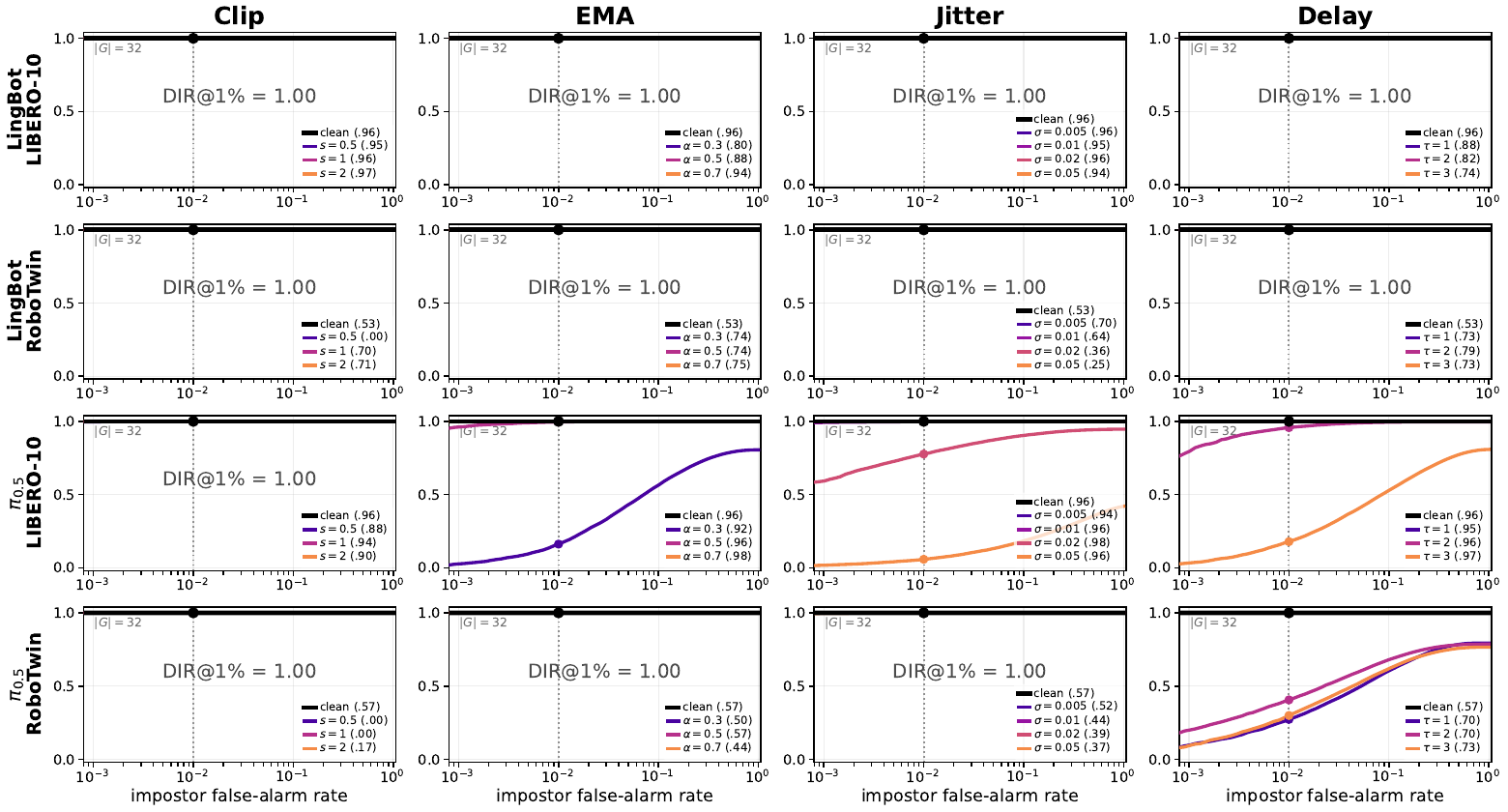}
\caption{Open-set identification under output-side attacks at group budget $|G|{=}32$. Each panel plots DIR against impostor false-alarm rate for one model--suite cell. Black is the no-attack condition; purple$\to$orange denotes weaker$\to$stronger attack settings. The dotted vertical line marks the $1\%$ FAR operating point, and parentheses give the fingerprinted policy's task success rate. Higher curves are better.}
\label{fig:identification-openset}
\end{figure*}
\mypara{Decision rules}
Given a probe group $G$, the verifier ranks all gallery keys by $T_G(k)$.
In the closed-set setting, the suspect is assumed to carry one gallery key, so the verifier returns the highest-scoring key $\hat k=\arg\max_k T_G(k)$.
We report the cumulative match characteristic (CMC): rank-$r$ is the probability that the true key $k^*$ appears among the top $r$ keys.
In the open-set setting, the suspect may carry no gallery key.
The verifier therefore returns the highest-scoring key only when $\max_k T_G(k)$ exceeds a rejection threshold $\tau_{\mathrm{id}}$; otherwise it abstains.
We set $\tau_{\mathrm{id}}$ from plain no-injection rollouts treated as impostors, using the $(1{-}\alpha_{\mathrm{FAR}})$ quantile of their $\max_k T_G(k)$ scores.
We report DIR@$\alpha_{\mathrm{FAR}}$, the fraction of genuine probes that both name the true key at rank $1$ and exceed $\tau_{\mathrm{id}}$.
Probe groups are bootstrap-resampled exactly as in \Cref{sec:eval-verification}.

\mypara{Clean identification}
\Cref{tab:identification-clean} shows that naming the key is usually not the limiting step once the fingerprint is detectable.
Identification is stricter than binary verification because the true key must outscore all $32$ decoys, not merely exceed one decision threshold.
The LingBot-VA cells clear this bar from one episode (rank-1 $=1.00$ on both LIBERO-10 and RoboTwin).
$\pi_{0.5}$ is harder from one episode (rank-1 $0.44$ on LIBERO-10 and $0.23$ on RoboTwin), matching the partial-observation and short-episode regime that also weakens its binary detection.
Aggregation closes the gap: by $|G|{=}16$, all four model--suite cells reach rank-1 $=1.00$ and open-set DIR@$1\%\geq0.99$.
The full clean CMC curves are in \Cref{app:identification-curves}.

\begin{table}[t]
\centering
\caption{Clean key identification over the $33$-key gallery. R1 and R5 are closed-set CMC at ranks $1$ and $5$ for group budgets $|G|{=}1$ and $|G|{=}16$. DIR is open-set identification at $|G|{=}16$, with thresholds set from plain no-injection impostor rollouts at FAR $1\%$ and $10\%$.}
\label{tab:identification-clean}
\small
\setlength{\tabcolsep}{3.5pt}
\resizebox{0.9\linewidth}{!}{\begin{tabular}{ll cc cc cc}
\toprule
& & \multicolumn{2}{c}{R1} & \multicolumn{2}{c}{R5} & \multicolumn{2}{c}{DIR} \\
\cmidrule(lr){3-4}\cmidrule(lr){5-6}\cmidrule(lr){7-8}
Model & Dataset & $G{=}1$ & $G{=}16$ & $G{=}1$ & $G{=}16$ & $1\%$ & $10\%$ \\
\midrule
LingBot & LIBERO-10 & 1.00 & 1.00 & 1.00 & 1.00 & 1.00 & 1.00 \\
LingBot & RoboTwin & 1.00 & 1.00 & 1.00 & 1.00 & 1.00 & 1.00 \\
$\pi_{0.5}$ & LIBERO-10 & 0.44 & 1.00 & 0.76 & 1.00 & 1.00 & 1.00 \\
$\pi_{0.5}$ & RoboTwin & 0.23 & 1.00 & 0.64 & 1.00 & 0.99 & 1.00 \\
\bottomrule
\end{tabular}
}\end{table}

\begin{table}[t]
\centering
\caption{Identification on owner-side variants over the $33$-key gallery. R1$_1$ is closed-set rank-1 at $|G|{=}1$; R1$_{16}$/DIR$_{16}$ and R1$_{64}$/DIR$_{64}$ report closed-set rank-1 and open-set detection-and-identification at $|G|{=}16$ and $64$. DIR uses FAR$=1\%$. }
\label{tab:identification-owner}
\small
\setlength{\tabcolsep}{3.5pt}
\resizebox{0.9\linewidth}{!}{\begin{tabular}{ll cc cc c}
\toprule
Variant & Model/Dataset & R1$_1$ & R1$_{16}$ & DIR$_{16}$ & R1$_{64}$ & DIR$_{64}$ \\
\midrule
LoRA & LingBot/LIBERO-10 & 1.00 & 1.00 & 1.00 & 1.00 & 1.00 \\
LoRA & LingBot/RoboTwin & 0.33 & 1.00 & 0.98 & 1.00 & 1.00 \\
LoRA & $\pi_{0.5}$/LIBERO-10 & 1.00 & 1.00 & 1.00 & 1.00 & 1.00 \\
LoRA & $\pi_{0.5}$/RoboTwin & 0.33 & 1.00 & 0.99 & 1.00 & 1.00 \\
\midrule
\texttt{prune30} & LingBot/LIBERO-10 & 1.00 & 1.00 & 1.00 & 1.00 & 1.00 \\
\texttt{prune30} & LingBot/RoboTwin & 1.00 & 1.00 & 1.00 & 1.00 & 1.00 \\
\texttt{prune30} & $\pi_{0.5}$/LIBERO-10 & 0.76 & 1.00 & 1.00 & 1.00 & 1.00 \\
\texttt{prune30} & $\pi_{0.5}$/RoboTwin & 0.07 & 0.54 & 0.03 & 0.96 & 0.44 \\
\midrule
\texttt{int8} & LingBot/LIBERO-10 & 1.00 & 1.00 & 1.00 & 1.00 & 1.00 \\
\texttt{int8} & LingBot/RoboTwin & 0.99 & 1.00 & 1.00 & 1.00 & 1.00 \\
\texttt{int8} & $\pi_{0.5}$/LIBERO-10 & 1.00 & 1.00 & 1.00 & 1.00 & 1.00 \\
\texttt{int8} & $\pi_{0.5}$/RoboTwin & 0.09 & 0.88 & 0.34 & 1.00 & 1.00 \\
\bottomrule
\end{tabular}
}\end{table}

\mypara{Output-side attacks}
When an output-side attack lowers the recovered true-key score, identification degrades in the same way binary detection does.
\Cref{fig:identification-openset} shows the open-set stress test at $|G|{=}32$.
Identification stays near the ceiling for both LingBot-VA cells and for most $\pi_{0.5}$ settings.
The visible drops occur in the same hard cases as binary detection: strong smoothing, jitter, or long delay attenuates the recovered signal before the gallery ranking step.
Thus the attacks do not create a new failure specific to key naming; they make the true key harder to separate from decoys.
The corresponding closed-set CMC curves are shown in \Cref{app:identification-robustness}.

\mypara{Owner-side variants}
Owner-side variants test whether identification still works when the observed actions come from a released descendant rather than the owner's base policy.
\Cref{tab:identification-owner} shows that most descendants become fully identifiable after aggregation.
LoRA descendants reach rank-1 $=1.00$ at $|G|{=}16$ in every model--suite cell, with DIR@$1\%\geq0.98$.
The LingBot-VA compression rows and the $\pi_{0.5}$/LIBERO-10 compression rows are also saturated at $|G|{=}16$.
The only weaker owner-side case is $\pi_{0.5}$/RoboTwin under action-side compression.
Even there, closed-set rank-1 rises to $0.96$ for \texttt{prune30} and $1.00$ for \texttt{int8} by $|G|{=}64$, although the open-set rejection threshold remains stricter for \texttt{prune30}.
This supports the same conclusion as the clean setting: once the fingerprint score is separable, naming the carried key is usually not the bottleneck.

\section{Security Analysis}
\label{sec:security}

The evaluation above shows that the owner's key produces a high group score $T_G(k^*)$.
Security analysis asks two further questions.
First, can some other key also produce a high score on the same suspect policy?
Second, can a white-box adversary edit the suspect policy so that the owner's key no longer scores highly?
We first analyze false-key risk.
This covers both independent owners whose assigned keys accidentally clear the threshold and adversaries who try many random keys until one appears to match the suspect.
Both cases are controlled by the false-key tail: the probability that a key $k'\ne k^*$ reaches the threshold.

For a decision threshold $\tau$, define the false-key collision tail and the true-key detection power as
\begin{align}
p_{\mathrm{coll}}(\tau)
&=
\Pr_{k'\ne k^*}\!\left[T_G(k')\ge \tau\right],
&
q_{\mathrm{true}}(\tau)
&=
\Pr\!\left[T_G(k^*)\ge \tau\right].
\label{eq:security-tail}
\end{align}
The verifier wants $p_{\mathrm{coll}}(\tau)$ small while keeping $q_{\mathrm{true}}(\tau)$ near one.
For $N$ owners with independently assigned $b$-bit keys, the chance that any other owner can pass the threshold is bounded by
\begin{equation}
P_{\mathrm{owner}}(N;\tau,b)
\le
N\,p_{\mathrm{coll}}(\tau)
+
\binom{N}{2}2^{-b}.
\label{eq:owner-population}
\end{equation}
The first term is the statistical false-key tail across $N$ owners; the second is the ordinary birthday collision probability for assigned keys.
For an adversary that tries $M$ independent random keys, the random-forgery probability and expected first-hit budget are
\begin{equation}
\begin{aligned}
P_{\mathrm{forge}}(M;\tau)
&=
1-\left(1-p_{\mathrm{coll}}(\tau)\right)^M,\\
\mathbb{E}[M_{\mathrm{first}}]
&\approx
\frac{1}{p_{\mathrm{coll}}(\tau)}.
\end{aligned}
\label{eq:forge-budget}
\end{equation}
We estimate these quantities with the same partial-observation MAP verifier used in the main experiments, with the matched filter and $|G|{=}16$.
The directly measured tail is an empirical operating point for this verifier and rollout pool, not a cryptographic proof.
When we report lower false-key rates, we label them as parametric operating points from the fitted tail model.

\mypara{Key Uniqueness}
Key uniqueness asks whether a different owner's assigned key could pass the same provenance test.
We estimate this risk by sampling $1024$ independent decoy keys over the $300$ fingerprinted episodes, calibrating each episode as in \Cref{eq:zscore}, and bootstrapping $2\times10^5$ groups of size $16$.
Each decoy key is treated as a possible independent-owner key.

\begin{figure}[t]
\centering
\includegraphics[width=\linewidth]{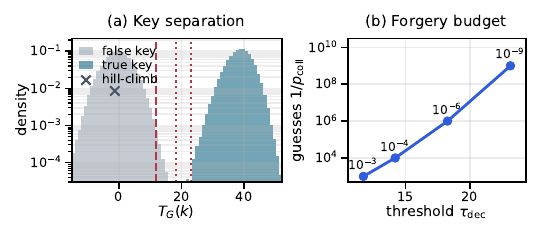}
\caption{Security operating points for $\pi_{0.5}$/LIBERO-10 with partial-observation MAP recovery and $|G|{=}16$. Left: bootstrap distributions of the group statistic $T_G$ for false keys and for the true key; vertical lines mark the empirical $10^{-3}$ threshold and the fitted Gaussian-tail thresholds for $10^{-6}$ and $10^{-9}$. Right: each threshold converted to the expected number of independent random-key guesses before one false key passes. }
\label{fig:security-summary}
\end{figure}

\Cref{fig:security-summary}(a) shows the main result: the false-key distribution is far left of the true-key distribution.
At the empirical $10^{-3}$ false-key operating point, $\tau_{\mathrm{dec}}{=}11.8$.
The false-key estimates agree across three views of the same tail: the empirical collision rate is $10^{-3}$ by construction, a Gaussian fit gives $9.4\times10^{-4}$ at the same threshold, and a GEV fit gives $1.1\times10^{-3}$.
The true key is far to the right, with $T_G(k^*)\sim\mathcal{N}(38.6,3.5^2)$ and $\mathrm{TPR}{=}1.00$ at $\tau_{\mathrm{dec}}{=}11.8$.

This separation lets the verifier consider lower tolerated false-key tails without losing the true key in the fitted model.
Under the Gaussian tail approximation, $\tau_{\mathrm{dec}}{=}18.3$ corresponds to $p_{\mathrm{coll}}\approx10^{-6}$ and $\tau_{\mathrm{dec}}{=}23.1$ corresponds to $p_{\mathrm{coll}}\approx10^{-9}$; both keep fitted true-positive rate $\mathrm{TPR}{=}1.00$ in this pool.
We treat these $10^{-6}$ and $10^{-9}$ values as model-based sensitivity points, not as distribution-free empirical bounds.
For owner populations, \Cref{eq:owner-population} separates false-key statistics from key assignment itself.
With $b{=}128$, the birthday term is about $1.5\times10^{-27}$ for $N{=}10^6$ owners.
The practical control is therefore the threshold: choose $\tau_{\mathrm{dec}}$ so that $N p_{\mathrm{coll}}(\tau_{\mathrm{dec}})$ is below the desired false-owner budget.

\mypara{Empirical Unforgeability}
Empirical unforgeability asks how many keys an adversary must try before one passes the verifier.
If the adversary knows the algorithm and verifier but not $k^*$, each independent random key succeeds with probability $p_{\mathrm{coll}}(\tau_{\mathrm{dec}})$.
At the empirical $10^{-3}$ operating point this gives an expected random-forgery budget of about $10^3$ guesses.
The fitted $10^{-6}$ and $10^{-9}$ operating points would raise that budget to about $10^6$ or $10^9$ guesses while preserving fitted true-positive rate $\mathrm{TPR}{=}1.00$ in \Cref{fig:security-summary}(b).
The deployed $128$-bit key space is the entropy ceiling, not the operating bottleneck.

A structured attack would need more than random trials: it would need a way to move through key space toward larger $T_G$.
The keyed reference is generated coordinate by coordinate as
\begin{equation}
\left(r_{k,\nu,c}\right)_d
=
G\!\left(\textsc{blake2b}(k\Vert\nu\Vert c\Vert d)\right),
\label{eq:keyed-reference-hash}
\end{equation}
where $d$ indexes the reference coordinate and $G$ expands the hash output into one Gaussian value for nonce $\nu$ and chunk $c$.
Thus the key reaches the score only through a one-way, non-differentiable seed map.
Empirically, nearby and offset keys are uncorrelated with the true reference ($|\rho|\le0.15$ for $k^*\!\pm\!\{1,2,3\}$, $k^*\!+\!100$, and $k^*\!+\!12345$).
A relaxed hill-climb over integer keys finds a lucky single-episode score, but it does not transfer across nonces: its group score is $T_G{=}{-}1.3$, far below $\tau_{\mathrm{dec}}$, while the true key scores $39.0$ on the same episodes.

These results support empirical unforgeability at the measured $10^{-3}$ operating point and show how the fitted tail model scales to stricter thresholds.
They do not prove key-extraction hardness in the cryptographic sense.
They show that, for this verifier and rollout pool, a forgery behaves like random guessing against a threshold that the verifier can tune using \Cref{eq:security-tail}.
\begin{figure}[t]
\centering
\includegraphics[width=0.9\columnwidth]{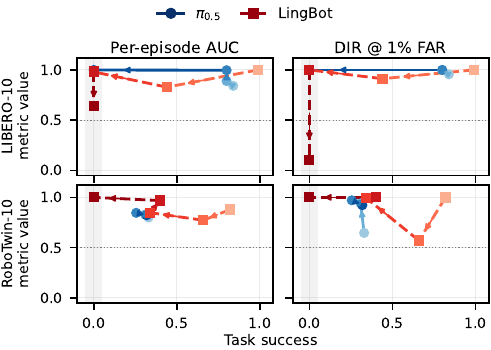}
\caption{Cost-utility view of direct adaptive removal.  Circles/solid lines are $\pi_{0.5}$; squares/dashed lines are LingBot-VA. Within each family, darker markers and arrowheads mean stronger adaptive attacks. A successful removal attack would appear in the lower-right region.}
\label{fig:attackd-cost-utility}
\end{figure}

\mypara{Adaptive Removal}
Adaptive removal tests the second security question: whether an adversary can edit the suspect policy to suppress the owner's key.
We give the adversary the suspect weights and the verifier, then let it fine-tune LoRA parameters $\theta'$ against the verifier.
The attack minimizes
\begin{equation}
\begin{aligned}
\mathcal{L}_{\mathrm{AD}}(\theta')
&=
\lambda_{\mathrm{rec}}\,
\ell_{\mathrm{rec}}(\hat z_{\theta'}, r_k)
+
\lambda_{\mathrm{task}}\,
\|a_{\theta'}-a_{\theta}\|_2^2 \\
&\quad+
\lambda_{\mathrm{reg}}\,
\|\theta'-\theta\|_2^2 .
\end{aligned}
\label{eq:attackd-loss}
\end{equation}
Here $\hat z_{\theta'}$ is the seed recovered by the same MAP verifier from the edited policy's action $a_{\theta'}$, $r_k$ is a keyed reference, and $\ell_{\mathrm{rec}}$ is the recovered-key score that the attacker tries to drive down.
The second term keeps the edited policy close to the pre-attack policy, and sweeping $\lambda_{\mathrm{task}}$ gives the cost--utility frontier.

This is a white-box stress test, stronger than the output-side attacks in \Cref{sec:eval-robust} and outside our default black-box audit setting.
It also tests the main removal concern: a strong single-rollout score gives the optimizer a local target, while a diffuse fingerprint gives the attacker little to suppress in any one rollout.

\Cref{fig:attackd-cost-utility} is read as a removal plot: a successful adaptive attack would appear in the lower-right, where task success stays high but the verification metric is low.
Across the two $\pi_{0.5}$ cells, there are no lower-right points.
LingBot-VA/LIBERO-10 also has no effective lower-right point: when DIR drops, task success has already fallen to $44\%$ or below.
LingBot-VA/RoboTwin-10 contributes the single borderline point, and only on DIR.
Relative to the no-attack anchor (DIR $0.999$ at $82\%$ task success), the highest task-preservation adaptive setting lowers DIR to $0.570$ while retaining $66\%$ task success; group detection AUC, however, remains $0.999$.
We therefore separate the claims: detection robustness holds in all four model--suite cells, while full multi-key identification robustness is supported in three cells and remains qualified for LingBot-VA/RoboTwin-10.
Details are in \Cref{app:adaptive-removal-details}.

\section{Discussion}
\label{sec:discussion-distillation}
\mypara{Distillation}
Distillation measures the far end of the evasion cost curve in \Cref{sec:threat-model}.
Here the adversary stops forwarding the protected service and trains a fresh student policy on outputs from the fingerprinted teacher.
The student may imitate the teacher's actions, but the adversary has now paid for a new artifact and runs it with a new sampler.
The deployed fingerprint preserves the sampler's noise law exactly: $z_c^{\mathrm{fp}}\sim\mathcal{N}(0,I_{D_z})$ whenever $z_c\sim\mathcal{N}(0,I_{D_z})$.
The action stream therefore does not contain a persistent pattern for behavior cloning to copy (\Cref{sec:injection,sec:eval-design-injection}).

\begin{figure}[t]
\centering
\includegraphics[width=\columnwidth]{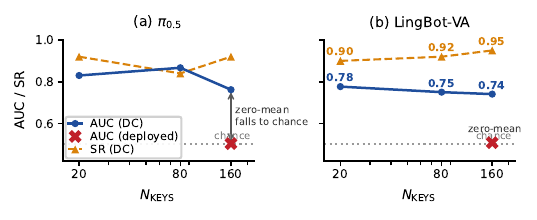}
\caption{Distillation stress tests. For biased seed-control marks, blue gives cross-student detection AUC and orange gives student task success (SR) as the lookup table varies over $N{=}20,80,160$ seed-bias vectors. Red crosses mark the deployed zero-mean seed fingerprint at comparable cardinality. Dashed gray lines mark chance AUC.}
\label{fig:distillation}
\end{figure}

\Cref{fig:distillation} compares the deployed high-entropy mark with a low-entropy mark designed to be easier for a student to learn.
The deployed zero-mean fingerprint falls to chance after behavior cloning on both $\pi_{0.5}$ and LingBot-VA (AUC $0.504$ and $0.508$).
For comparison, we also test a small lookup table of seed-bias vectors.
The current robot state selects one bias vector before each chunk, which creates a repeatable action pattern that a student can learn.
This biased seed-control mark survives behavior cloning better than the deployed mark (AUC $0.74$--$0.87$ across the two panels) because it gives the student a persistent pattern to learn.
That persistence comes from giving up exact Gaussian noise-law invariance and the large key space used for unforgeability and multi-key deployment.
The experiment therefore frames distillation resistance as a trade-off between inheritance through behavior cloning and hidden, high-entropy keyed provenance.
The stress-test setup is in \Cref{app:distillation-details}.

\section{Conclusion}
\label{sec:conclusion}

We presented keyed latent-provenance verification for deployed VLA and world-action policies.
The method watermarks the generator's latent seed, keeps the sampler noise law unchanged, and verifies ownership from the executed action channels available in a black-box audit.
Our evaluation shows that this signal supports ownership detection and multi-key identification across model families, robot embodiments, output-side attacks, and owner-side variants.
These results make latent-seed provenance a practical protection mechanism for private generative robot policies.

\section*{Ethics Considerations}
This work develops a defensive capability: a way for the legitimate owner of a robot policy to verify, after the fact, whether a suspect deployment was derived from their model. We considered the parties the method touches: the model owner it protects, the adversary it is designed to detect, the operators and bystanders of a fingerprinted robot, and independent third parties who could be wrongly implicated. Two of these concerns need a direct answer: that verification could wrongly implicate an innocent party, and that the technique could be repurposed beyond provenance.

\mypara{Bounding false attribution} A provenance tool is harmful if it can be used to implicate an innocent party, and our security analysis addresses this directly. Uniqueness shows that an independent owner's key does not trip another owner's verifier, so a clean, unrelated policy is not falsely claimed as the owner's. Unforgeability shows that an adversary who knows the algorithm and the verifier code, but not the secret key, cannot fabricate a key that passes. A third party therefore cannot be framed by a planted mark. Verification also requires the owner's secret key together with authorized access to the suspect, for example through legal or regulatory inspection. It cannot be run covertly against an arbitrary party.

\mypara{Dual-use} Like any watermark, the technique could in principle be repurposed, for instance to tag a model for tracking. Two properties intrinsic to the design limit this risk. The mark is keyed and recoverable only by the holder of the secret key. Verification is an authorized, after-the-fact audit rather than a passive broadcast signal. The benefit is that owners of expensive, easily-copied robot policies gain a way to demonstrate provenance. We judge this to outweigh the residual risk, which the keyed, authorized-audit design already limits. We are also explicit that the guarantees are empirical operating points, not cryptographic ones. The method is therefore not a basis for over-reliance in adversarial or legal settings.

\section*{Generative AI Usage Considerations}
We did not use generative AI in this work, except for grammar checking and light rephrasing of text the authors had already written. It played no part in ideation, literature search or data analysis. The generative models named in the paper, the VLA policy and the LingBot world-action model, are the objects of our study, not authoring aids.

\bibliographystyle{plainurl}
\bibliography{ref}

\appendix
\subsection{Method Hyperparameters}
\label{app:method-hyperparameters}
All main experiments use latent injection strength $\beta=1.0$.
In \Cref{eq:injection}, any $\beta \in [0,1]$ preserves the marginal Gaussian law of the sampler noise, while larger $\beta$ gives the verifier a stronger correlation with the keyed reference.
We therefore use the largest value for the deployed latent fingerprint.
The output-space baseline in \Cref{sec:eval-design-injection} uses its own action-amplitude scale and is not part of this latent-$\beta$ choice.

The selector uses a per-episode cap of $m=5$ selected chunks.
The gap parameter $P$ controls how densely those selected chunks can appear.
We use $P=1$ for the $\pi_{0.5}$ cells, $P=6$ for LingBot-VA/LIBERO-10, and $P=2$ for LingBot-VA/RoboTwin.
The smaller LingBot-VA/RoboTwin value is used because these dual-arm episodes are shorter; a larger gap can leave some episodes with no selected chunk.
If an episode ends before all $m$ selected positions occur, the verifier scores the selected chunks that were actually generated.

For MAP recovery, all cells use prior weight $\lambda_z=1.0$.
The observation scale $\sigma_{\mathrm{obs}}$ is fixed by model implementation rather than tuned per attack: $\sigma_{\mathrm{obs}}=10^{-4}$ for $\pi_{0.5}$ and $\sigma_{\mathrm{obs}}=10^{-3}$ for LingBot-VA.
This scale sets the optimizer's relative weight on action fitting, while decoy-key calibration absorbs model- and task-level score-scale differences before aggregation.

\subsection{Distillation Stress-Test Details}
\label{app:distillation-details}
For $\pi_{0.5}$/LIBERO-10, each distillation arm relabels the demonstration corpus with teacher-generated actions, keeping the observations fixed.
We then train an attention-only LoRA student for $1500$ steps and roll it out without the owner's keyed sampler.
The deployed-fingerprint arm uses the zero-mean seed reference from the main method.
The diagnostic arms replace it with a small lookup table of $N{=}20,80,160$ seed-bias vectors.
At each chunk, we quantize the current robot state, hash it, and use the hash modulo $N$ to select which bias vector to add to the starting noise.
Detection uses the base-model MAP verifier with the true key and $32$ decoys, and cross-student AUC compares the marked-student scores to a clean-student null.

For LingBot-VA, we repeat the biased seed-control sweep on LIBERO-10 with $10$ tasks and $10$ episodes per task.
The students are again evaluated without injection and scored against a clean student with the same verifier structure.

\subsection{Adaptive-Removal Stress-Test Details}
\label{app:adaptive-removal-details}
The adaptive-removal stress test instantiates \Cref{eq:attackd-loss} with LoRA edits.
We fix the recovery-loss weight and sweep the task-preservation weight, producing the cost--utility frontier in \Cref{fig:attackd-cost-utility}.
After fine-tuning, we evaluate each edited policy with the same decoy-key calibration and group aggregation used in the main experiments.

For $\pi_{0.5}$/LIBERO-10, this stress test covers a non-adaptive anchor and four adversarial settings.
Across the sweep, the fingerprint remains detectable at the standard group budget: TPR at $|G|{=}16$ stays at $0.99998$ or higher.
The strongest setting collapses task success for both watermarked and plain rollouts to zero, while detection remains at the ceiling.
Thus, in the diffuse-score regime, the white-box optimizer does not find a usable removal direction; it can only damage the policy.

For LingBot-VA/LIBERO-10, the non-adaptive anchor is already saturated under the same verifier.
The reason to test LingBot-VA separately is mechanistic rather than procedural: its stronger single-rollout seed recovery gives an adaptive optimizer a cleaner target than $\pi_{0.5}$.
At the strongest setting, task success collapses to zero and group identification falls to $0.103$, which is a utility failure rather than a useful removal attack.

The RoboTwin cells extend the same stress test to the dual-arm setting.
For $\pi_{0.5}$, neither detection nor identification moves to the lower-right region.
For LingBot-VA, the no-attack anchor has DIR $0.999$ at $82\%$ task success.
The closest adaptive point lowers DIR to $0.570$ while keeping $66\%$ task success, so it is a real reduction in multi-key identification margin.
It is not a detection removal point: group AUC remains $0.999$, and the other adaptive settings either recover near-ceiling identification or collapse task success to zero.
This cell is therefore a limitation for strict identification, not for binary ownership detection.

\subsection{Budget Sweeps for Weak Cells}
\label{app:budget-curve}
\begin{figure}[t]
\centering
\includegraphics[width=\columnwidth]{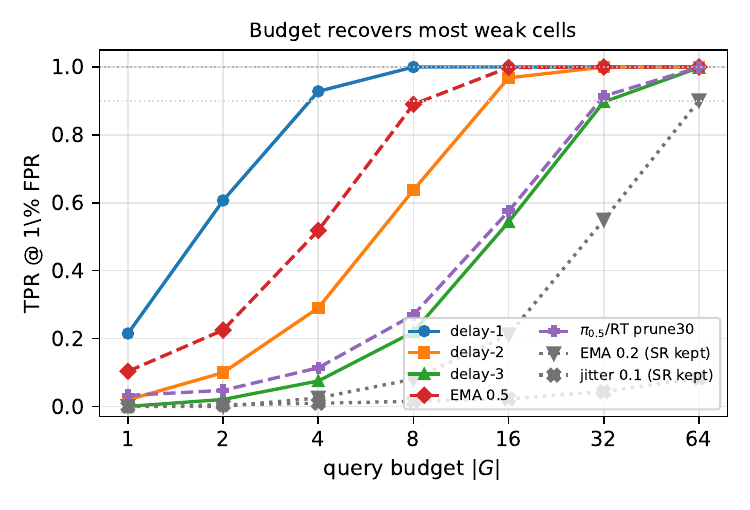}
\caption{Detection rate versus query budget for $\pi_{0.5}$ cells that are not saturated at small budgets. Each line is TPR at $1\%$ FPR as the group size $|G|$ grows on a log scale; higher and further left is better. Delay uses the verifier's global synchronization search. The delayed cells, default smoothing, and RoboTwin pruning recover with a larger budget. The strongest jitter setting remains near the floor, marking the boundary where output perturbations preserve task success but suppress detection.}
\label{fig:budget-curve}
\end{figure}

The main text summarizes the query-budget effect for cells that remain below ceiling at $|G|{=}16$.
\Cref{fig:budget-curve} plots the budget sweep for these weak cells, and \Cref{tab:robust-detection} gives the full robustness table behind the compact main-text summary.

\begin{table*}[t]\centering
\caption{Detection robustness across the $2{\times}2$ family design under adversary output-removal attacks and owner-side weight variants. Each cell gives per-episode AUC (AUC$_1$, $|G|{=}1$, fingerprinted vs.\ plain), then AUC and TPR at FPR$=1\%$ for group budgets $|G|{=}16$ and $|G|{=}64$ (the vertical rule separates AUC from TPR within each budget). For \texttt{clip}/\texttt{EMA}/\texttt{jitter}, lo/$\star$/hi denote the smallest, canonical, and largest swept strength; \texttt{delay} is per step. The verifier uses synchronization search for constant delay and otherwise estimates $\ell^*{=}0$. `--' marks a condition absent for that family. $^\dagger$For $\pi_{0.5}$/RoboTwin the compression rows restrict the edit to the action expert (${\sim}16\%$ of parameters), since compressing the whole model discards the vision--language backbone.}
\label{tab:robust-detection}\small
\setlength{\tabcolsep}{3pt}
\resizebox{\textwidth}{!}{%
\begin{tabular}{l c c|c c|c c c|c c|c c c|c c|c c c|c c|c}
\toprule
 & \multicolumn{5}{c}{LingBot/LIBERO-10} & \multicolumn{5}{c}{LingBot/RoboTwin} & \multicolumn{5}{c}{$\pi_{0.5}$/LIBERO-10} & \multicolumn{5}{c}{$\pi_{0.5}$/RoboTwin$^\dagger$} \\
\cmidrule(lr){2-6}\cmidrule(lr){7-11}\cmidrule(lr){12-16}\cmidrule(lr){17-21}
 & \multicolumn{1}{c}{} & \multicolumn{2}{c}{$|G|{=}16$} & \multicolumn{2}{c}{$|G|{=}64$} & \multicolumn{1}{c}{} & \multicolumn{2}{c}{$|G|{=}16$} & \multicolumn{2}{c}{$|G|{=}64$} & \multicolumn{1}{c}{} & \multicolumn{2}{c}{$|G|{=}16$} & \multicolumn{2}{c}{$|G|{=}64$} & \multicolumn{1}{c}{} & \multicolumn{2}{c}{$|G|{=}16$} & \multicolumn{2}{c}{$|G|{=}64$} \\
\cmidrule(lr){3-4}\cmidrule(lr){5-6}\cmidrule(lr){8-9}\cmidrule(lr){10-11}\cmidrule(lr){13-14}\cmidrule(lr){15-16}\cmidrule(lr){18-19}\cmidrule(lr){20-21}
Condition & AUC$_1$ & AUC & TPR & AUC & TPR & AUC$_1$ & AUC & TPR & AUC & TPR & AUC$_1$ & AUC & TPR & AUC & TPR & AUC$_1$ & AUC & TPR & AUC & TPR \\
\midrule
none (clean) & 1.00 & 1.00 & 1.00 & 1.00 & 1.00 & 1.00 & 1.00 & 1.00 & 1.00 & 1.00 & 0.90 & 1.00 & 1.00 & 1.00 & 1.00 & 0.85 & 1.00 & 1.00 & 1.00 & 1.00 \\
\midrule
\multicolumn{21}{l}{\emph{Output removal attacks}} \\
\texttt{clip}\,lo & 1.00 & 1.00 & 1.00 & 1.00 & 1.00 & 1.00 & 1.00 & 1.00 & 1.00 & 1.00 & 0.64 & 0.93 & 0.39 & 1.00 & 0.97 & 0.86 & 1.00 & 1.00 & 1.00 & 1.00 \\
\texttt{clip}\,$\star$ & 1.00 & 1.00 & 1.00 & 1.00 & 1.00 & 1.00 & 1.00 & 1.00 & 1.00 & 1.00 & 0.97 & 1.00 & 1.00 & 1.00 & 1.00 & 0.85 & 1.00 & 1.00 & 1.00 & 1.00 \\
\texttt{clip}\,hi & 1.00 & 1.00 & 1.00 & 1.00 & 1.00 & 1.00 & 1.00 & 1.00 & 1.00 & 1.00 & 0.97 & 1.00 & 1.00 & 1.00 & 1.00 & 0.87 & 1.00 & 1.00 & 1.00 & 1.00 \\
\texttt{EMA}\,lo & 1.00 & 1.00 & 1.00 & 1.00 & 1.00 & 1.00 & 1.00 & 1.00 & 1.00 & 1.00 & 0.54 & 0.67 & 0.02 & 0.81 & 0.09 & 0.86 & 1.00 & 1.00 & 1.00 & 1.00 \\
\texttt{EMA}\,$\star$ & 1.00 & 1.00 & 1.00 & 1.00 & 1.00 & 1.00 & 1.00 & 1.00 & 1.00 & 1.00 & 0.70 & 0.98 & 0.69 & 1.00 & 1.00 & 0.86 & 1.00 & 1.00 & 1.00 & 1.00 \\
\texttt{EMA}\,hi & 1.00 & 1.00 & 1.00 & 1.00 & 1.00 & 1.00 & 1.00 & 1.00 & 1.00 & 1.00 & 0.89 & 1.00 & 1.00 & 1.00 & 1.00 & 0.86 & 1.00 & 1.00 & 1.00 & 1.00 \\
\texttt{jitter}\,lo & 1.00 & 1.00 & 1.00 & 1.00 & 1.00 & 1.00 & 1.00 & 1.00 & 1.00 & 1.00 & 0.84 & 1.00 & 1.00 & 1.00 & 1.00 & 0.88 & 1.00 & 1.00 & 1.00 & 1.00 \\
\texttt{jitter}\,$\star$ & 1.00 & 1.00 & 1.00 & 1.00 & 1.00 & 1.00 & 1.00 & 1.00 & 1.00 & 1.00 & 0.73 & 0.99 & 0.85 & 1.00 & 1.00 & 0.85 & 1.00 & 1.00 & 1.00 & 1.00 \\
\texttt{jitter}\,hi & 1.00 & 1.00 & 1.00 & 1.00 & 1.00 & 1.00 & 1.00 & 1.00 & 1.00 & 1.00 & 0.56 & 0.70 & 0.05 & 0.85 & 0.17 & 0.81 & 1.00 & 1.00 & 1.00 & 1.00 \\
\texttt{delay} 1 & 1.00 & 1.00 & 1.00 & 1.00 & 1.00 & 1.00 & 1.00 & 1.00 & 1.00 & 1.00 & 0.81 & 1.00 & 1.00 & 1.00 & 1.00 & 0.71 & 0.99 & 0.84 & 1.00 & 1.00 \\
\texttt{delay} 2 & 1.00 & 1.00 & 1.00 & 1.00 & 1.00 & 1.00 & 1.00 & 1.00 & 1.00 & 1.00 & 0.68 & 0.97 & 0.41 & 1.00 & 1.00 & 0.71 & 0.99 & 0.84 & 1.00 & 1.00 \\
\texttt{delay} 3 & 1.00 & 1.00 & 1.00 & 1.00 & 1.00 & 0.99 & 1.00 & 1.00 & 1.00 & 1.00 & 0.65 & 0.94 & 0.26 & 1.00 & 0.98 & 0.71 & 0.99 & 0.83 & 1.00 & 1.00 \\
\midrule
\multicolumn{21}{l}{\emph{Owner-side weight variants}} \\
LoRA finetune & 1.00 & 1.00 & 1.00 & 1.00 & 1.00 & 0.83 & 1.00 & 1.00 & 1.00 & 1.00 & 1.00 & 1.00 & 1.00 & 1.00 & 1.00 & 0.87 & 1.00 & 1.00 & 1.00 & 1.00 \\
\texttt{prune30} & 1.00 & 1.00 & 1.00 & 1.00 & 1.00 & 1.00 & 1.00 & 1.00 & 1.00 & 1.00 & 0.99 & 1.00 & 1.00 & 1.00 & 1.00 & 0.68 & 0.97 & 0.56 & 1.00 & 1.00 \\
\texttt{int8} & 1.00 & 1.00 & 1.00 & 1.00 & 1.00 & 1.00 & 1.00 & 1.00 & 1.00 & 1.00 & 1.00 & 1.00 & 1.00 & 1.00 & 1.00 & 0.76 & 1.00 & 0.94 & 1.00 & 1.00 \\
\bottomrule
\end{tabular}
}
\end{table*}

\subsection{Synchronization-Search Ablation}
\label{app:lagsearch}
This ablation isolates the alignment step on the hardest delayed cell, $\pi_{0.5}$/LIBERO-10.
\Cref{tab:lagsearch} shows that, at $|G|{=}16$, one-step delay recovers from TPR $0.28$ to $1.00$ when the search is enabled.
Longer delays remain weak under lag-zero scoring, but the global search realigns them: two-step delay is nearly saturated by $|G|{=}16$, and three-step delay reaches the ceiling by $|G|{=}64$.
\begin{table}[h]\centering
\caption{Synchronization-search ablation on the \texttt{delay} attack, $\pi_{0.5}$/LIBERO-10. Rows are delay length; columns sweep the query budget $|G|$. Each entry is TPR at $1\%$ FPR (AUC in the $|G|{=}1$ column), with the lag search off (lag-zero) and on (the estimated global $\tau^*$). A constant delay is a pure time shift the global search marginalizes, so $\tau^*$ recovers it; the residual at longer delay is graceful attenuation that aggregation closes. $\pi_{0.5}$/RoboTwin and every non-delay condition estimate $\tau^*{=}0$ and are unchanged by the search (omitted).}
\label{tab:lagsearch}\small
\resizebox{\linewidth}{!}{\begin{tabular}{c cccc c cccc}
\toprule
 & \multicolumn{4}{c}{lag-zero} & \multicolumn{4}{c}{global $\tau^*$ search} \\
\cmidrule(lr){2-5}\cmidrule(lr){6-9}
delay & AUC & \multicolumn{3}{c}{TPR @ $|G|$} & AUC & \multicolumn{3}{c}{TPR @ $|G|$} \\
 & ($|G|{=}1$) & 16 & 32 & 64 & ($|G|{=}1$) & 16 & 32 & 64 \\
\midrule
1 & 0.63 & 0.28 & 0.62 & 0.95 & 0.91 & 1.00 & 1.00 & 1.00 \\
2 & 0.54 & 0.02 & 0.04 & 0.09 & 0.77 & 0.96 & 1.00 & 1.00 \\
3 & 0.48 & 0.00 & 0.00 & 0.00 & 0.68 & 0.53 & 0.90 & 1.00 \\
\bottomrule
\end{tabular}
}\end{table}

\subsection{Same-Task vs.\ Cross-Task Aggregation}
\label{app:aggregation-mode}
\begin{figure}[t]
\centering
\includegraphics[width=\columnwidth]{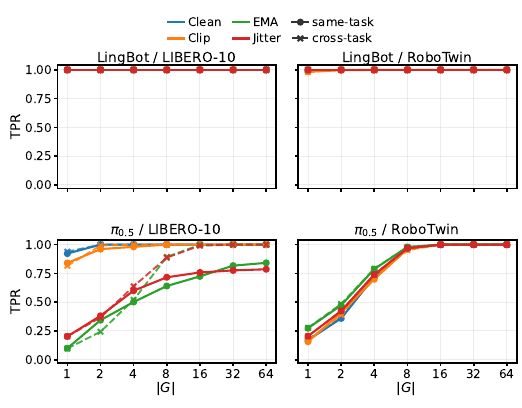}
\caption{Same-task vs.\ cross-task aggregation at FPR$=1\%$. Solid curves restrict each group to one task; dashed curves group rollouts across tasks. The $\pi_{0.5}$ panels are shown for completeness but are not a meaningful same-task comparison: $\pi_{0.5}$/LIBERO-10 has one episode per task, and $\pi_{0.5}$/RoboTwin has a single evaluation task.}
\label{fig:aggregation-mode}
\end{figure}

\Cref{prop:rate} assumes that grouped episodes are roughly independent, while the audit setting allows the verifier to collect rollouts on either the same task or different tasks.
The meaningful comparison is on the LingBot-VA cells, where there are enough rollouts per task for a same-task bootstrap.
\Cref{fig:aggregation-mode} shows that the same-task and cross-task curves nearly overlap there, so the aggregation gain is not an artifact of mixing tasks.

\subsection{Detailed Clean Identification Curves}
\label{app:identification-curves}
\begin{figure}[t]
\centering
\includegraphics[width=\columnwidth]{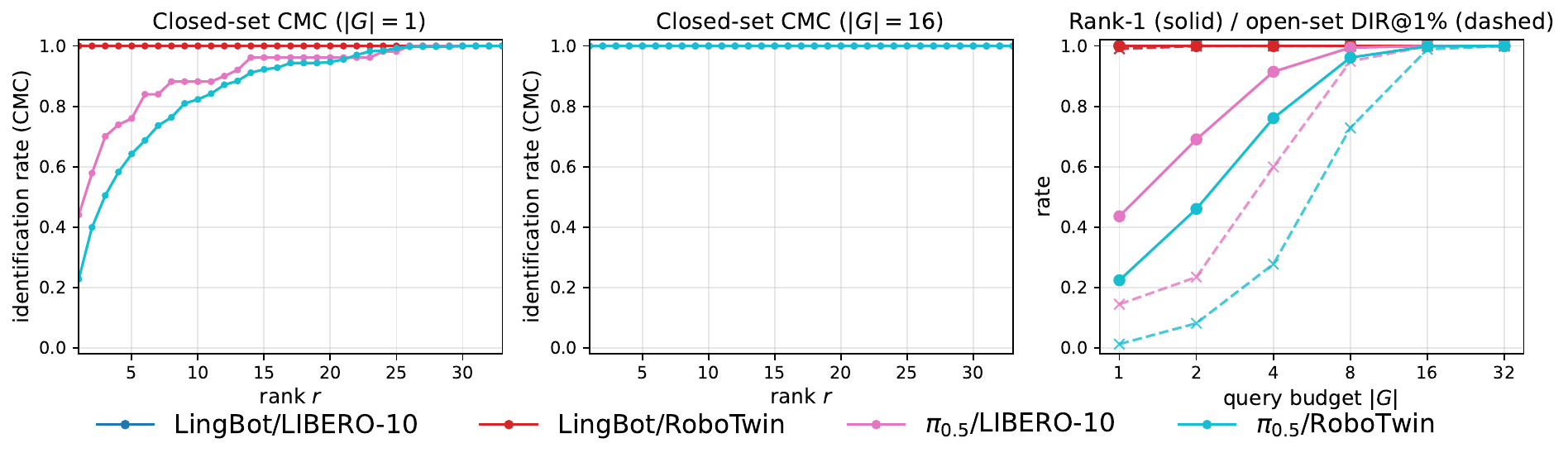}
\caption{Clean-policy identification curves over the $33$-key gallery. Left and middle: CMC curves at $|G|{=}1$ and $|G|{=}16$, where rank-$r$ is the probability that $k^*$ is among the top $r$ keys. Right: closed-set rank-1 and open-set DIR@$1\%$ FAR as the query budget grows.}
\label{fig:identification}
\end{figure}

\Cref{fig:identification} expands the clean identification result into CMC curves and budget-dependent rank-1/open-set identification rates.

\subsection{Detailed Identification Robustness}
\label{app:identification-robustness}
\begin{figure*}[t]
\centering
\includegraphics[width=\textwidth]{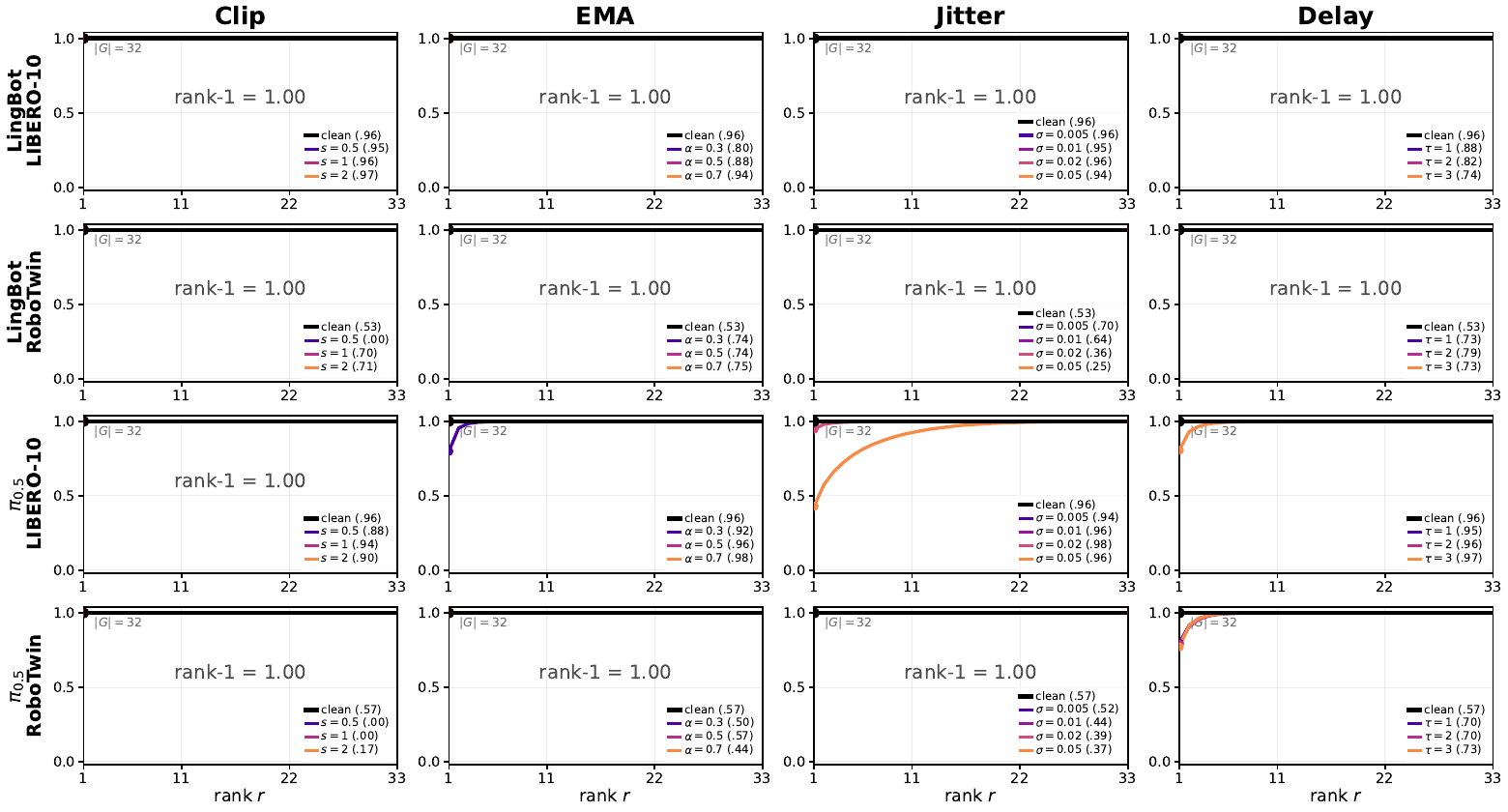}
\caption{Closed-set identification robustness: cumulative match characteristic (CMC) over the $33$-key gallery under output-side attacks at group budget $|G|{=}32$. Rows are model--suite cells; columns are attack families; black is the clean policy and purple$\to$orange means weaker$\to$stronger attack. Each curve is $P(\text{rank of }k^* \le r)$ vs.\ rank $r$; the dot at rank $1$ is rank-1 identification.}
\label{fig:identification-combined}
\end{figure*}

\Cref{fig:identification-combined} reports the full closed-set CMC curves under output-side attacks.

\end{document}